\documentclass{aa}
\usepackage{indentfirst}
\usepackage{natbib}
\usepackage{graphicx}
\usepackage{amsmath,amssymb}

\newcommand{\zp}{$z_{\rm ph}$}
\newcommand{\be}[1]{\begin{equation}\label{#1}}
\newcommand{\ee}{\end{equation}}
\newcommand{\reffig}[1]{Fig~\ref{#1}}
\newcommand{\der}{\,{\rm d}}
\newcommand{\ddroit}[2]{\tfrac{\der #1}{\der #2}}

\newcommand{\hmkpc}{\,h^{-1}\,{\rm kpc}}
\DeclareMathOperator{\hypergeom}{F}
\newcommand{\hypgeo}{\sideset{_2}{_1}\hypergeom}

\begin{document}
\title{A radial mass profile analysis of the lensing cluster MS2137.3-2353.
  \thanks{Based on observations obtained at the Very Large Telescope (VLT) at Cerro Paranal operated by European Southern Observatory.} }

\author{R. Gavazzi\inst{1} \and B. Fort\inst{1} \and Y. Mellier\inst{1,2}
  \and R. Pell\'o\inst{3} \and  M. Dantel-Fort\inst{2} }
\institute{Institut d'Astrophysique de Paris, UMR 7095, 98 bis
  Boulevard Arago, 75014 Paris, France \and Observatoire de Paris,
  LERMA, 61 Av. de l'Observatoire, 75014 Paris, France \and
  Observatoire Midi-Pyr\'{e}n\'{e}es, UMR 5572, 14 Av. E. Belin,
  31400 Toulouse, France}
\date{Received xxx / Accepted xxx}
\offprints{(Rapha\"el Gavazzi) gavazzi@iap.fr}

\abstract{We reanalyze the strong lens modeling of the cluster of galaxies
  MS2137.3-2353 using a new $UBVRIJK$ data set obtained with the ESO Very Large
  Telescope. We infer the photometric redshifts of the two main arc systems
  which are both found to be at $z=1.6\pm 0.1$. After subtraction of the
  central cD star light in the previous F702/HST imaging we found only one
  object lying underneath. This object has the expected properties
  of the fifth image associated to the tangential arc.
  It lies at the right location, shows the right orientation and
  has the expected signal-to-noise ratio.
  
   We improve the previous lens modelings of the central dark matter
  distribution of the cluster, using two density profiles:
  an isothermal model with a core, and the NFW-like model with a cusp.
  Without the fifth image, the arc properties together with the shear map 
  profile are equally well fit by  the isothermal model and  by a sub-class 
  of generalized-NFW mass profiles having 
  inner slope power index in the range $0.7\leq \alpha \leq 1.2$. Adding
  new constrains on the center lens position provided by the fifth
  image favors isothermal profiles that better predict the fifth image 
  properties.
   A detailed model including nearby cluster galaxy perturbations or the
  effect of the stellar mass distribution to the total mass inward
  does not change our conclusions but imposes the $M/L_I$ of the cD
  stellar component is below 10 at a 99\% confidence level.

  Using our new detailed strong+weak lensing model together with Chandra X-ray
  data and the cD stellar component we finally discuss intrinsic properties
  of the gravitational potential. Whereas X-ray and dark matter have a
  similar orientation and ellipticity at various radius,
  the cD stellar isophotes are twisted by $13^\circ \pm 3^\circ$.
  The sub-arc-second  azimuthal shift we observe between the
  radial arc position and the predictions of elliptical models
  correspond to what is expected from a mass distribution twist.
  This shift may result from a projection effect of the cD and the cluster
  halos, thus revealing the triaxiality of the mass components.
}

\titlerunning{Radial mass profile of MS2137.3-2353}
\maketitle

\section{Introduction}
Cosmological N-body simulations of hierarchical structures formation in a
universe dominated by collision-less dark matter predict universal density
profiles of halos that can be approximated by the following distribution
\be{nfw}
\rho(r)=\rho_s \left(\tfrac{r}{r_s}\right)^{-\alpha} \left[1+\tfrac{r}{r_s}\right]^{\alpha-3}.
\ee
The early simulations of \citep{NFW} (hereafter NFW) found $\alpha = 1$,
leading to profiles with a central {\sl cusp} $\alpha$ and
an asymptotic $r^{-3}$ slope, steeper than isothermal (hereafter IS).

More recently, simulations with higher mass resolution confirmed
that the density profile Eq. \eqref{nfw} can fit the dark matter distribution
of halos, although different values of $\alpha$ were obtained by various
authors \citep[see {\sl e.g.~}][]{Ghigna00,Bullock01}.

While the collision-less $\Lambda CDM$ cosmology explains observations
of the universe on large scales, two issues concerning these
halos are still debated. The first one is the apparent excess of sub-halos
predicted in numerical simulations, compared to the number of satellites
in halos around normal galaxies \citep{Klypin99, Moore99}.
This discrepancy may be resolved if some of the sub-halos never
formed stars in the past and are therefore dark structures
 \citep{BKW01,Verde02}. \citet{Metc01,Keea,Keec} or
\citet{Dalal02} argued that we may already see effects of such dark
halos through the perturbations they induce on the magnification on the
gravitational pairs of distant QSOs.\\
The second prediction is the existence of a cuspy
universal profile which cannot explain the rotation curves of
dwarf galaxies \citep{Saluccietal00}. If these discrepancies
are not simply due to a resolution problem of numerical
simulations, then, as it was pointed out by several
authors, they may illustrate a small-scale crisis for current CDM
models \citep{NavSt}. In order to solve these issues, alternatives
to pure collision-less cold dark matter particles,
have been proposed \citep{Spergel00,bode01}. Also several
physical mechanisms which could change the inner slope of mass profiles,
like central super-massive black holes \citep{Merritt,Haehnelt02},
tidal-merging processes inward massive halos \citep{Dekel02} or
adiabatic compression of dark matter can be advocated
\citep[see {\sl e.g.}~][]{Blumenthal,Keea}.

The demonstration that halos do follow a NFW mass profile over a wide
range of mass scale would therefore be a very strong argument in favor
of collision-less dark matter particles. Unfortunately, and despite
important efforts, there is still no conclusive evidence that
observations single out the universal NFW-likes profile and
rule out other models. Clusters of galaxies studies are among the most
puzzling. In general, weak lensing analysis or X-rays emission models show that
both singular isothermal sphere (SIS) and NFW fit equally well their
dark matter profile, but there are still contradictory results which
seem to rule out either NFW or IS models
\citep[see for example][]{Allen98, Tysonetal98, Mellier99, Cloweetal00,
 Willick00, Cloweetal01, Arabadjis02, Athreya02}.
This degeneracy is explained  because most observations probe
the density profile at intermediate radial distances, where an IS
and a NFW profiles have a similar $r^{-2}$ behavior.

A promising attempt to address the cusp-core debate is to model
gravitational lenses with multiple arcs which are spread at different
radial distances, where the SIS and the NFW slopes may differ
significantly. As emphasized by \citet{Miralda95},
ideal configurations are clusters with a simple geometrical
structure (no clumps) and with the measurements of the stellar
velocity dispersion profile of its central galaxy \citep[See e.g.][]{Kelson02}.
The  MS2137.3-2353 cluster satisfies these requirements and turns out to
be an exceptional lensing configuration with several lensed images,
including a demagnified one we find out in this work at the very center
of the lensing potential. In this paper, we analyze the possibility
to break the degeneracy between IS and NFW mass profiles using new
data set of MS2137.3-2353 obtained at the VLT and the properties of
this new fifth image.

The paper is organized as follows. In Sect. \ref{observ_section} we review
the cluster properties after a summary on previous modelings that claimed
for very deep photometric observations. This section also presents the new VLT
observations and describes the optical properties of the cluster.
Sect. \ref{modeling_section} presents the strong lensing models for
softened IS elliptical halos and NFW cuspy profiles. We discuss
the global agreement of both approaches within the CDM paradigm in Sect.
\ref{discus_section}. We stress the importance of the detection of the fifth
central demagnified image of the tangential arc system and discuss the
observational prospects for the near future in Sect. \ref{conclu_section}.
Throughout this paper, we assume a $\Omega_0 = 0.3,\;\Omega_\Lambda = 0.7$,
and $ H_0 \equiv 100\;h\;{\rm km}\:{\rm s}^{-1}\:{\rm kpc}^{-1}$ cosmology
in which case $ 1 \arcsec = 3.24\hmkpc$ at the cluster redshift $z=0.313$.

\section{The MS2137.3-2353 lens configuration}\label{observ_section}
\subsection{Overview}
MS2137.3-2353 is a rich cD cluster of galaxies located at $z_l=0.313$
\citep{Stocke91}.
The central region ($ \lesssim 4 \arcmin$) does not show any substructures
and has a regular visible appearance, as expected for a well dynamically
relaxed gravitational system. The discovery of a double arc configuration,
among which was the first radial arc \citep{Fort92}, makes MS2137.3-2353 a
perfect cluster for modeling, without the need for complex mass distribution.

\begin{figure*}[htp]
  \centering
  \resizebox{!}{17cm}{\includegraphics{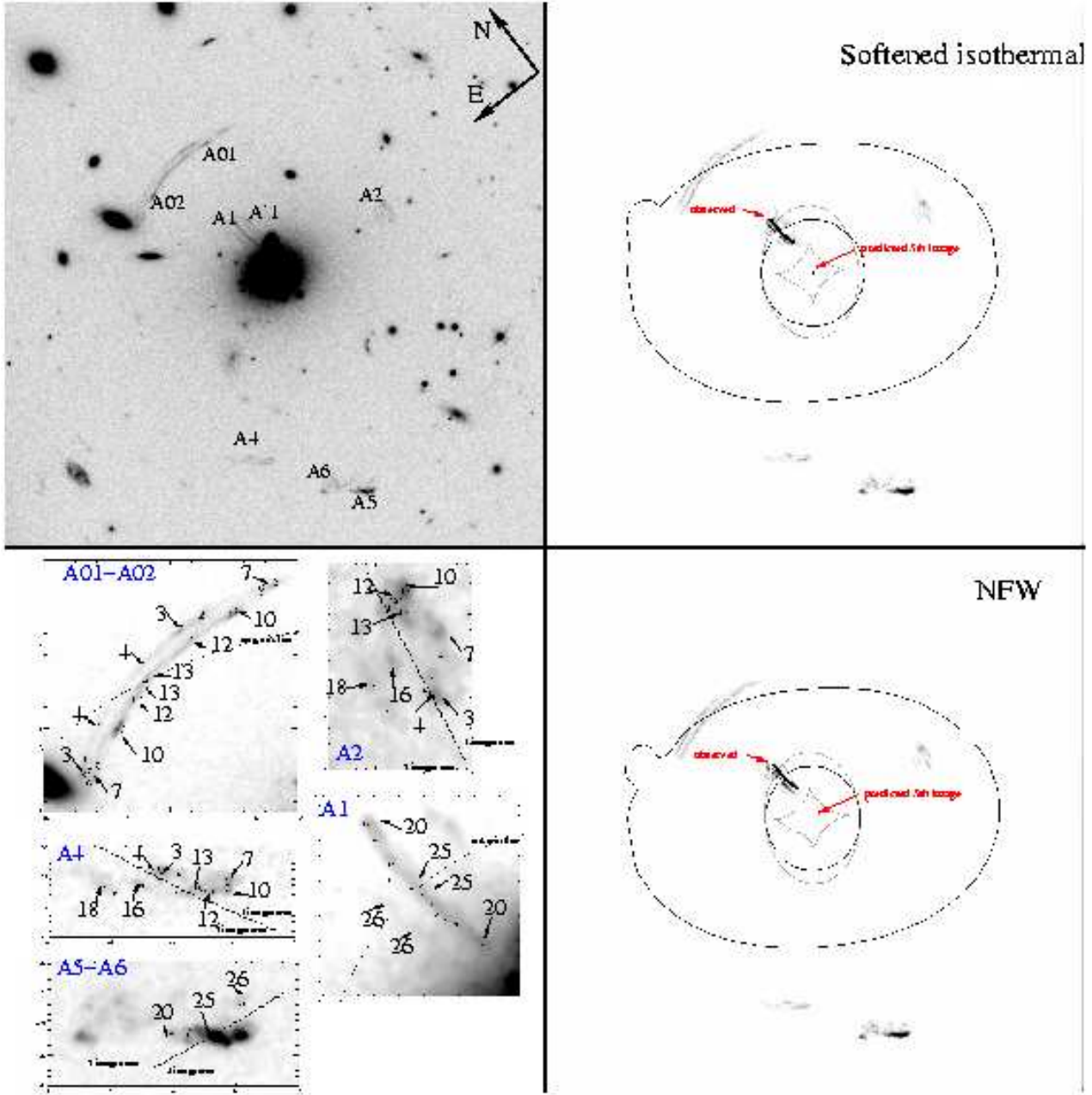}}
  \caption{Upper left panel : overview of the lens configuration.
    The three arcs systems \{A01,A02,A2,A4\}, \{A1,A5\} and \{A'1,A6\}. The
    central cD galaxy. This F702 HST field is $56\times56$ arcsec wide
    ({\sl i.e.} $180\times 180\,\hmkpc$). Upper (resp. lower) right
    panel : reconstruction of arcs deduced from the single component best
    fit IS (resp. NFW) model (see \ref{model_1c_subsec}).
    In these panels are reported the observed radial arc location.
    The small azimuthal offset is discussed in Sect. \ref{azimprof_subsec}.
    The fifth demagnified image predicted by the models near the
    center is detailed in \reffig{center_fig}. Lower left panel,
    detail of some dots used for the model fitting (see Table 
    \ref{point_table_append}).}
  \label{genima_fig}
\end{figure*}

The lens generates a tangential arc (A01-A02, see \reffig{genima_fig})
associated with two other counter-images A2 and A4 positioned
around the cD galaxy. A01 and A02 are twin images with reverse parity.
They are two merging ``partial'' images of the
source element located inside the tangential caustic line.
The lens potential is expected to produce a fifth demagnified image near
the center, but the cD galaxy brightness peak hampers its direct detection.
In Sect. \ref{5th_cent_observ_subsec}, we investigate in more details the
presence of a candidate and the detection probability of this fiducial image.

The lens also gives rise to a radial arc A1 partially buried beneath the
stellar diffuse component of the cD. This arc is associated with
the elongated image A5. \citet{Hammer97} argued the diffuse object A6
near A5 is probably  another counter-arc associated with the diffuse light
A'1 which encompasses A1. The lens configuration is shown in
\reffig{genima_fig}. The radial arc at about 5 arcsec together with the
tangential one at 16 arc-sec already probe the potential at two different
radii and provide a unique way to determine its slope in this region.
Furthermore, a radial arc together with its counter-image gives a
stronger constraint than a tangential system on the potential ellipticity.

\subsection{Previous lens models of MS2137.3-2353}
This ideal configuration has early prompted \citet{Mellier93} and
\citet{Miralda95} to show that an isothermal elliptical model with a
small core radius ($r_c < 30 \hmkpc$ ) remarkably well reproduces the
gravitational images pattern.

Thanks to the high spatial resolution of HST images, \citet{Hammer97} were
able to confirm the lens configuration described by \citeauthor{Mellier93}
and to better constrain the location and the shape of the counter image of
the radial arc. They derived the properties of the mass distribution,
assuming a $\beta$-model
\begin{equation}\label{hamprof_eq}
  \rho(r) = \rho_0 \left(1+(r/r_c)^2\right)^{-\beta},
\end{equation}
with $ \beta = 0.87 \pm 0.04$, $r_c = 2.25 \pm 0.75 \arcsec$.
This model confirmed that arc properties observed in lensing
clusters dominated by giant elliptical galaxies can be interpreted
with potential well centered on their brightest cluster members.
This trend is indeed robust enough to be generalized with a fair
confidence level on similar clusters. Hence, only small
deviations around central galaxy positions may eventually be explored.

For all these models the average orientation and ellipticity of
the potential are kept unchanged with radius and match the stellar
light halos of the cD galaxy.
\citet{Miralda95} studied the dynamical state of the central stellar
halo and predicted their radial velocity dispersion profile.
Similar studies were carried out on several clusters of
galaxies where a tight correlation is found between the projected
dark matter (DM) distribution and the faint stars halo \citep{Kneib93,Kneib96}.
Later, \citet{Miralda02} argued that the large tangential
deviation angle between the radial image of MS2137.3-2353 and its opposite
counter image implies the dark matter distribution to have a large
ellipticity. It is worth noticing that self-interacting dark matter
models predict central halos must be circular ; so \citeauthor{Miralda02}'s
argument may rule out these particles.

Regarding its radial dark matter profile, despite the tight constraint
provided by the radial arc on isothermal models with core, alternative mass
profiles can naturally explain its properties. \citet{Bartel96}
demonstrated that the radial arc in MS2137.3-2353 is also consistent with a
NFW profile. It can easily produce models as good as isothermal spheres with
core radius making the radial arc properties of MS2137.3-2353 less useful
than previously expected. A primary problem was the complete ignorance
of the arc redshifts. Models just predicted that the radial and tangential
arcs could be at almost the same redshift, if below $z \approx 1$,
or both at a large redshift. However, any conclusions on the inner slope
of the potential are sensitive to these redshifts.

Besides, in order to probe cuspy profiles one need to explore
the innermost region of the lens, where a 5th demagnified image
associated to a fold arc system is expected to form.
This task requires a careful galaxy subtraction and an accurate lens
model which can predict whether the differences between the 5th
image properties between a NFW profile and an isothermal sphere
are significant and measurable. These goals were serious limitations to
previous modelings that could use for high resolution imagery. Fortunately,
they are no longer restrictions when the recent observations by Chandra
\citep{wise01} and by the VLT (this work) are used together with HST data.
The new constraints provided by these new data sets on the geometry of the
baryonic and non-baryonic matter components and on the lensed images
properties permit for the first time to probe the mass profile of a cluster
over three decades in radius, {\sl i.e.} from 1 kpc up to 1 Mpc.

\subsection{New insight on the light distribution}\label{data_section}
The HST data have been obtained from the Space Telescope archive.
They consist in 10 WFPC2 images obtained with the F702W
filter\footnote{Program ID: 5402; PI: Gioia}.
The individual frames were stacked using the IRAF/STSDAS package,
leading to a final exposure time of 22,000 sec. In addition,
we used new data sets obtained during Summer 2001 with
the VLT/FORS instrument in optical $UVI$ bands and with the VLT/ISAAC
instrument in $J$ and $K$\footnote{Program ID: 67.A-0098(A) FORS and
  67.A-0098(B) ISAAC; PI: Mellier}.
The FORS and ISAAC data have been processed at the TERAPIX data
center\footnote{{\tt  http://terapix.iap.fr}}.
Pre-calibrations, astrometric and photometric calibrations as well
as image stacking were done using standard CCD image processing
algorithms. We also  used the $B$ and $R$ images kindly provided by
S. Seitz that were obtained by the FORS team during the
1999 and 2000 periods. The exposure times of these data are shorter
than our $UVI$ and $JK$ data, but they are still useful for the
photometric redshift estimates.

\begin{table}[h]
  \centering\label{VLT_photom}
  \caption{A brief summary of the VLT data.
    The first column summarizes the seeing of the final stacked images,
    the total exposure time is given in the second column. Also given
    the Magnitude Zero Points (Z.P.). The $B$ and $R$ images were obtained
    by the FORS team (provided by S. Seitz).}
  \begin{tabular}{cccc}\hline\hline
    Filter&Seeing($\arcsec$)& Exp. time (sec)& Z.P (mag) \\ \hline
    U & 0.72 & 5280 & 30.856 \\
    B & 1.2  & 2400 & 32.888 \\
    V & 0.64 & 6900 & 33.978 \\
    R & 0.58 &  300 & 32.501 \\
    I & 0.69 & 12000 & 33.484 \\
    J & 0.49 & 5880 & 27.643 \\
    K & 0.50 & 6480 & 26.797\\\hline
  \end{tabular}
\end{table}

\begin{figure}[h]
  \resizebox{\hsize}{!}{\includegraphics{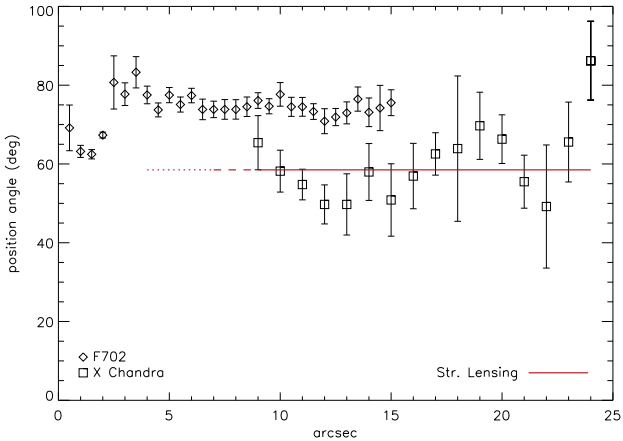}}
  \caption{Orientation of the isophotal major axis as a function of radius.
    ({\sl squares:} intra-cluster gas, {\sl diamonds:} stars in
    the F702 band). The horizontal line represents the average orientation
    of the DM halo from modeling beyond $\sim8\arcsec$.
    See Sect. \ref{modeling_section}.}
  \label{baryons_fig}
\end{figure}

The MS2137.3-2353 optical data provide the azimuthal stellar light
distribution and show that its geometry is elliptical.
Its ellipticity\footnote{All ellipticities discussed here are defined 
  as $\epsilon={a^2-b^2 \over a^2+b^2}$, where $a$ and $b$ are the major
  and minor axes.} increases with radius, starting from an almost
circular shape at the center, and reaches quickly a constant value of
0.30 beyond the giant tangential arc location ($r \ge 15 \arcsec$).
The position angle is $PA \approx (71\pm 4)^\circ$ at $r = 15 \arcsec$
(see \reffig{baryons_fig}).
Assuming a fiducial mass-to-light ratio $\Upsilon_I=2$ and a
I-band K-correction of 0.23, we evaluate the rest-frame I luminosity
$L_I = 1.9 \times 10^{11}\,h^{-1}\,{\rm L}_\odot$.

The early ROSAT results of \citet{Gioia90} and \citet{Ettori99} and the
recent Chandra observations of \citet{wise01} provide additional clues
on the cluster halo. They confirm it appears as a well relaxed cluster.
The X-isophotes are remarkably elliptical\footnote{ we used the task {\it ellipse}
  in the IRAF/STSDAS package for isophotal fitting.} and do not show substructures.
The orientation of gas is almost constant $PA_X = 58 ^\circ \pm 7^\circ$,
(see \reffig{baryons_fig}). A new interesting observational feature
is the global misalignment between the diffuse stellar component and the
hot intra-cluster gas. It suggests that the stellar light distribution
does not match exactly the DM distribution. This point is independently
confirmed by strong lensing models and is discussed in Sect.
\ref{azimprof_subsec}.
The MS2137.3-2353 radial properties inferred from X-rays data
reveal that the brightness profile presents a $r_c\sim7\arcsec$ core radius
and an asymptotic slope $\alpha \sim 1.17$, and an index $\beta \sim 0.56$.
\citet{Ettori99,Allen} modeled the X-ray emission and derived a gas mass
fraction $f_{rm gas}\approx 0.10-0.15$ depending on the inferred cosmology.
In both cases, this value is almost constant between 30 and $300\hmkpc$.
 Since the geometry of X-ray emission follows the overall potential and
represents a small and constant mass fraction,we will not consider
separately the gas and the dark matter in the lens modeling of MS2137.3-2353
in the following. Instead, we will simply reduce both components
to an effective dark matter halo as the sum of the gas and true DM model.
As a prospect, a good refinement would be the full introduction of this
component, independantly of the DM halo. Moreover, a fully 3D deprojected
modeling of both the X-ray emissivity and the strong lensing arcs system
would certainly be the next requirement for future modelings.

\subsection{VLT photometry \& redshifts determination}\label{redphot_subsec}
The photometric redshifts of arcs have been measured with the {\tt hyperz}
software \citep{Bolzonellaetal00,Pello01}. The redshift \zp~ is derived from
a comparison between the spectral energy distribution of galaxies inferred
from the $UBVRIJK$ photometry and a set of spectral templates of galaxies
which are followed with look-back time according to the evolution models
of \citet{Bruzual93} \citep[see][for~details]{Athreya02}. The validation of
{\tt hyperz} is discussed in \cite{Bolzonellaetal00} and has been already
validated using spectroscopic redshifts on many galaxy samples.
With the $UBVRIJK$ set of filters, it is possible to measure all
redshifts of our selected galaxy sample lying in the range $0.0<z<3.5$.
The expected redshift accuracy is between $\pm 0.05$ and  $\pm 0.2$,
depending on the magnitude of each arc, which is enough to scale the
convergence of a lens model.

For each arc, the $UBVRI$ and $JK$ photometry was done as follows.
We used {\tt SExtractor} \citep{Bertin96} to estimate magnitudes
in $2 \arcsec$ apertures around a well defined barycenter for each part of
the arcs. The $V$ frame is taken as the reference since arcs are
significantly bluer than the cD light. We also tried to take the $U$ and
$J$ ones to check the robustness of the method. As well, results are stable
against variations of aperture.

\begin{table}[ht]
  \centering \caption{Photometric redshifts. Uncertainties take into
    account the scatter in the best fits with different choices for
    photometric measurements (aperture size, reference filter...).
    Note that HST data are not used for photometry.}
  \begin{tabular}{lcccc}
    \hline \hline
    Arc &A01    &A02    &A2 &A4 \\\hline \zp
    &$1.64\pm0.15$&$1.6\pm0.1$&$1.7\pm0.2$&$1.58\pm0.2$\\\hline \hline
    Arc &A1 &A5 &A'1    &A6 \\\hline
    \zp &--&$1.60\pm0.1$&--&$1.1\pm0.3$\\\hline
  \end{tabular}\label{zphot_tab}
\end{table}
\begin{figure}[h]
  \resizebox{\hsize}{!}{\includegraphics{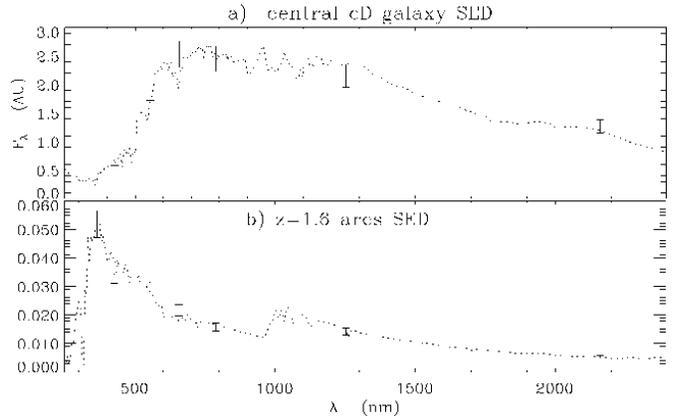}}
  \caption{Spectral Energy Distributions resulting from photometric redshift
    analysis. a) Central cD galaxy at z=0.313 and b) arc A5 deduced to be
    at $z\simeq 1.6$. One can see that contrast between arcs and cD is 15
    times higher in $U$ then in a redder filter like F702.}
  \label{SED_fig}
\end{figure}

For the radial arcs A1 and A'1, photometry is strongly sensitive to the
foreground cD diffuse stellar component. Furthermore, A1 and A'1 are
overlapping, so no estimation of photometric redshifts are really stable
for these objects. A better estimation of their redshift is provided by
their counter-arcs which both are free from contamination. Results for
all multiple images systems are summarized in table \ref{zphot_tab}.\\
Taking the best determination, we conclude that $z_s = 1.6 \pm 0.1$ for the
two sources responsible of the radial and tangential arc systems. The models
detailed in Sect. \ref{best1c_models_subsub} explain the need for a
different redshift of the source responsible of A'1 and A6 and is
consistent with the photometric redshift $z_{\rm s,A'1-A6} \approx 1.1$.
Hence, the critical density at the cluster redshift and with the adopted
cosmology is: $$\Sigma_{\rm crit}=3.73\times 10^{9}\,h\,{\rm M}_\odot
  {\rm kpc}^{-2}.$$
\citet{Sand} have recently reported a spectroscopic determination
of the redshift of arcs which are both found to be at $z = 1.501$ in
remarkably good agreement with our color determination.
From a gravitational lensing and mass estimate 
point of view, the difference between
their spectroscopic redshift and our photometric prediction is 
un-significant. The geometric efficiency term $D_{ds}/D_{s}$ is slowly 
varying at this
redshift. We nevertheless use their redshift estimation and the following
value for the critical density:
$$\Sigma_{\rm crit}=3.78\times 10^{9}\,h\,{\rm M}_\odot {\rm kpc}^{-2}.$$

\subsection{Detection of the fifth central image}\label{5th_cent_observ_subsec}
Gravitational optics  with a smooth potential and no central singularities
predict strong magnification should produce an odd number of lensed
images \citep{Burke81,SEF92}. More generally, the location, the
demagnification or even the lack of the central image are in principle
clues on the properties of the innermost density profile of lenses.

In the case of MS2137.3-2353, we expect the large arc A0 to have a
fifth demagnified counter-image. Unfortunately, any simple mass models
of the lens configuration predicts the fifth image of this fold configuration
should lie within  one or two arc-seconds from the cluster center,
that is inside the central cD light distribution. Its detection is
therefore uncertain and depends on its surface brightness, its size and
its color with respect to the cD light properties.

In order to check whether the fifth image associated to A0 is technically
detectable, we made several lens models using different mass profiles which
all successfully reproduce the tangential and radial arcs together with
their corresponding counter-images. We predict its position $\vec{r}_5$ and
magnification from the softened IS and the NFW profiles of Sect.
\ref{model_1c_subsec}. They are respectively $\mu_{\rm IS}(\vec{r}_5) = 0.2$ and
$\mu_{\rm NFW}(\vec{r}_5)=0.1$. The signal-to-noise ratio per HST/F702 pixel
yields :
\be{sn_br}
\left(\frac{S}{N}\right)_{\rm F702}^{\rm pix} =
\left(\frac{N_5}{1+N_{\rm cD}/N_5+N_{\rm sky}/N_5}\right)^{1/2} \simeq 0.7
\ee
where $N_5$, $N_{\rm cD}$ and $N_{\rm sky}$ are the number of photo-e$^-$ from
respectively the fifth image, the cD and the sky background close to
$\vec{r}_{5}$. Taking into account the size of the image, we can
express the signal-to-noise in terms of flux, $\left(S/N\right)_{\rm F702}^{F}$,
as a function of the magnification $\mu(\vec{r}_5)$ (assuming that the
magnification $\mu(\vec{r}_{\rm A2})$ does not much change with models).
\begin{equation}\label{sn_fl}
  \left(\frac{S}{N}\right)_{\rm F702}^{\rm F} \simeq 7 \times
  \sqrt{\mu(\vec{r}_5)}.
\end{equation}
The expected signal-to-noise ratio is $\approx$ 3 in the IS case and
$\approx$ 2 for the NFW profile. So, in principle, the fifth image of the
MS2137.3-2353 lensing configuration is detectable.

Using the counter-image A2 of area $A_{\rm A2}$ and flux $F_{\rm A2}$,
we reconstructed the predicted fifth image satisfying :
$$\left\{ \begin{array}{cc}
  F_{5} = \mu(\vec{r}_5) \times \mu(\vec{r}_{\rm A2})^{-1} \times F_{\rm A2} \\
  A_{5} = \mu(\vec{r}_5) \times \mu(\vec{r}_{\rm A2})^{-1} \times A_{\rm A2}
\end{array}\right.$$
and inserted it inside the cD galaxy at several positions close (but different)
to the expected location $\vec{r}_5$. We then determined the significance
of several extraction-detection techniques on the Space Telescope image.
A Mexican-hat compensated filter turned out to provide the best cD light
subtraction and an optimal detection of the fifth image twins we put
inside at different positions.
In all cases it was detected exactly at the right position, whatever its
location inside the cD and with the expected signal-to-noise.\\
Because we used a compensated filter which smoothes the signal,
this later is not straightforward and we had to compare the amplitude
of the flux contained in the extracted object to the variance of the
background contained inside independent cells of similar size ranging
along concentric annuli located at the radius where simulated fifth image
twins are putted ($0.6 \arcsec \lesssim r \lesssim 0.9 \arcsec$).
The averaged S/N found in annuli is 2.6, but it scatters between 1.3 and
3.5 depending on the local noise properties.

\begin{table}[h]
  \centering
  \caption{Properties of the fifth images on real data (R.) and
    predicted from the best lens modeling (IS or NFW). The positions
    $(x,y)$ are given in arc-sec, with respect to the cD centroid.
    Position angles (P.A.) are given in degrees and $a/b$ is the axis ratio.
    The errors are found from the changes when varying some {\tt SExtractor}
    parameters.
    Although both the position angles and the ellipticity of the IS
    and NFW are compatible with the data, there is a significant difference
    in positions. The offset $\vert\delta\vec{x}\vert$ between the IS and
    the real position is only $0.16\arcsec$, whereas it is $0.36\arcsec$
    for NFW, which is larger than uncertainties on observations
    (third column $\vert\delta \vec{x}\vert \sim 0.05\arcsec$)}
  \begin{small}\begin{tabular}{cccccc}\hline\hline
      ID &$(x,y)$& $\vert\delta\vec{x}\vert$ &P.A.& $a/b$ & S/N \\ \hline
      R. & (0.64;0.70)& 0.05 & $28\pm 14$ & $3.1\pm1.3$ & 2.6\\
      IS & (0.52;0.81)& 0.16 & 15.  & 2.2& 2.6\\
      NFW & (0.33;0.52)& 0.36 & 27. & 2.2& 2.1\\\hline
  \end{tabular}\end{small}\label{fifthtable}
\end{table}

\begin{figure*}[ht]
  \centering
  \resizebox{17cm}{!}{\includegraphics{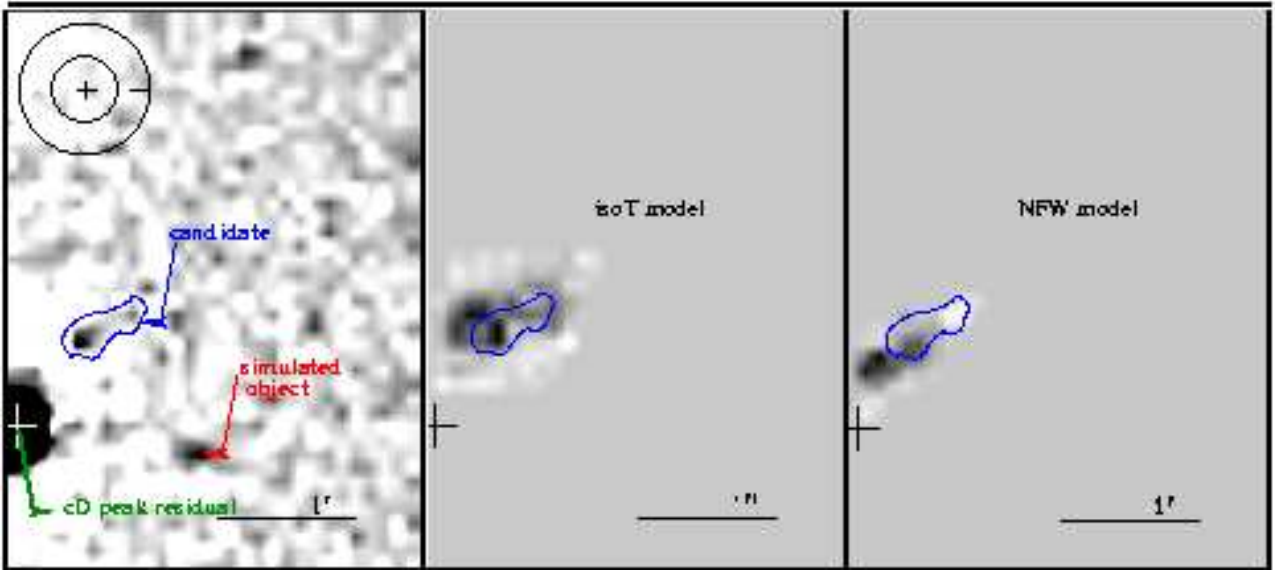}}
  \caption{Visualization of residuals from subtraction of the cD brightness.
    Orientation is the same as in \reffig{genima_fig}. Left panel: HST
    filtering. One can see the real residual as well as a simulated object
    derived from the IS model shown on the right panel. Note the comparable
    signal-to-noise. Also plotted the Mexican-hat filter size with
    both positive ($r\leq 1.5\, {\rm px}$) and negative areas ($1.5\, {\rm px}
    \leq r \leq 3.5\, {\rm px}$)
    Also shown the noise-free reconstructed 5th image of the fold arc expected
    from IS (middle panel) and NFW (right panel) strong lensing best fit
    models centered on the cD. The position is consistent with
    the IS model. Note that the filtering algorithm reduces the high
    frequencies in the upper frame. Thus, shape measurements just provide
    rough estimates of ellipticities, but accurate orientations. Note also 
    that the isophote plotted here, is illustrative and does not correspond 
    to the any quantitative shape measurement.}
  \label{center_fig}
\end{figure*}

The application of the extraction technique on the real data is
straightforward. The brightest residual in the filtered frame shown
in the right panel of \reffig{center_fig} is detected at the expected
location when compared to models and is clearly the most obvious object
underneath the cD. The object properties are listed in Table \ref{fifthtable}.
They are remarkably similar to the IS and NFW fifth image predictions.
Its coordinates are however closer to the IS fifth image than the NFW model.
The signal-to-noise ratio of the candidate is $\simeq 2.5$,
in very good agreement with our expectations. In the frame of \reffig{genima_fig},
the centroid position of the candidate is at
\be{pos_fifth_eq}
\vec{r}_5 = (0.64 \arcsec; 0.70\arcsec),\quad r_5 = 0.9\arcsec.
\ee
Despite its poorly resolved shape, the candidate exhibits an orientation
$PA \simeq 28\pm 14^\circ$ and an axis ratio $a/b = 3.1 \pm 1.4$,
in good agreement with the values predicted by both models
(see Table \ref{fifthtable}). It is worth noticing that even the
morphology of the fifth image shows similarities with the reconstructed images.
In particular, it shows a bright extension inward and a smaller faint spot
outward as if it would be dominated by two sub-clumps which are also visible
on predictions of Fig. \ref{center_fig}. 
In the following, when using the fifth image knowledge, we apply different
weights on the detected features for the lens modeling. The brightest part of 
the fifth image is statistically significant and is associated to the dot labeled
(10) on Fig. \ref{genima_fig} and table \ref{point_table_append}. The mappings
of the other labeled dots are not as significant in the fifth image. Thus, we apply
a $\sim 9$ times smaller weight ({\sl i.e.} 3 times larger errorbars). In other
words, the fifth image is almost reduced to a point-like information without 
shape measurements.

\section{The dark matter distribution in MS2137.3-2353}\label{modeling_section}
In this section, the properties of the dark matter distribution of
MS2137.3-2353 
are discussed in view of the most recent constraints we obtained from VLT data.
We first revisit a single potential model using only strong lensing data but
no fifth image. We then compare the projected mass profiles of the best NFW
and IS models, extrapolated beyond the giant arcs positions, with the
weak lensing analysis. Finally the fifth image is included in the strong
lensing model which is used together with the weak lensing and the cD stellar
halos to produce a comprehensive model of the different mass components.

\subsection{Strong lensing optimization method}\label{optim_subsec}
The optimization have been carried out with the {\tt lensmodel}
\footnote{{\tt http://astro.uchicago.edu/\~{}ckeeton/gravlens/}}
\citep{Keeb} inversion software. This alternative to the \citet{Mellier93}
or \citet{Kneib93,Kneib96} algorithms allows us to check the efficiency and
the accuracy of this software for arc modeling and to take advantage of
its association tool for multiple point-images. This facility was initially
developed by Keeton for multiple-QSOs but turns out to be well suited for
HST images of extended lensed objects. The images association is performed by
identifying conjugated substructures like bulges in extended images.
Because of the surface brightness conservation, brightest areas in an
image map into the brightest of the associated ones.

Our modeling started by identifying the brightest conjugate knots in
each image. More precisely, when the identification of $N_{\rm knots}$
distinct features in images is completed (with respectively
$N_{j=1\ldots N_{\rm knots}}$ multiplicity) one can write
$N_{\rm knots}$ times the lens equation relating source and image
positions and the lens potential $\phi$:
\be{lens_eq}
\vec{\theta}_{S_j}=\left\{\begin{array}{l}
 \vec{\theta}_1-\nabla\phi(\vec{\theta}_1)\\
 \ldots\ldots\\
 \vec{\theta}_{N_j}-\nabla\phi(\vec{\theta}_{N_j})\\
\end{array}\right.
\ee
This yields the following $\chi^2$ definition calculated in the image plane :
\begin{equation}\label{chisq_def_eq}
  \chi^2_{\rm img}=\sum_{j=1}^{N_{\rm knots}}\sum_{i=1}^{N_j}
  \delta\vec{x}_{ij}^T S^{-1}_{ij} \delta\vec{x}_{ij},
\end{equation}
with
\be{freedom_eq}
2\sum_{j=1}^{N_{\rm knots}}(N_j - 1) - N_{\rm par}
\ee
degrees of freedom, where $N_{\rm par}$ is the number of free parameters
in the model. Here, $S^{-1}_{ij}$ is the error matrix for the position of
knot $j$ in the image $i$ and 
$\delta\vec{x}_{ij}=\vec{x}_{{\rm obs},ij}-\vec{x}_{ {\rm mod},ij}$.
Analogous $\chi^2$ minimization can be done in the source plane in order
to speed up the convergence process.
It is only an approximation of the previous one that does not directly
handle observational errors in the image plane. However, it is much
faster because it does not need to invert Eq. \eqref{lens_eq}.
Once the minimum location is roughly found, one can use the image
plane $\chi^2_{\rm img}$ to determine the best parameter set with
a better accuracy.

It is worth noting that the uncertainties in the conjugate points positioning
done during the association process dominates the astrometric errors in
the position of each knot. Typically, the systematic uncertainty is of
order $0.1-0.2\arcsec$. The VLT color similarities were also used to confirm
the associations. The mapping between extended images is given by the
magnification matrix $a$ :
\begin{equation}\label{magnif_mat}
  a^{-1}_{ij}(\vec{\theta}) =
  \delta_{ij} - \partial_i\partial_j \phi(\vec{\theta}),
\end{equation}
where $\delta_{ij}$ is the Kronecker symbol.
Hence, other fainter conjugate dots become easier to identify once the local
linear transformation between multiple images is known. The procedure
``get constraints''-``fit a model'' can be iterated to use progressively
more and more informations.

In MS2137.3-2353, we kept 13 unambiguous quintuple conjugated dots in the
tangential arc A01. Each one is associated to four different dots in A02,
A2, A4 and the fifth image. We selected also 6 dots in the parts of A2 and
A4 that are only triply imaged. Likewise, A1 is decomposed in two symmetric
merging images and is also associated with the Eastern part of A5
(6 triple conjugated dots). Fig. \ref{genima_fig} and table 
\ref{point_table_append} summarize the associations we selected.

The various models are actually over-constrained. The 6 free parameters are
detailed in the following section. Following Eq. \eqref{freedom_eq},
the number of constraints is
\begin{small}
  \begin{equation}\label{constraints_eq}
    N = 2\times \{ 13 (5 - 2 ) + 6 ( 3 - 2) + 7 ( 3 - 1 ) \} =118.
  \end{equation}
\end{small}
The first term corresponds to the regions of the tangential system which are
imaged five times, whereas the second term refers to regions imaged three
times. The third term correspond to the radial system which is imaged three times.
\footnote{ Eq. \eqref{constraints_eq}: the $-2$ in parenthesis includes
  the unknown source position and the unused fifth central image.
  When using this later, the number of constraints is 156.}
Nevertheless, only 25 of these 118 constraints appear significant to
represent the first and second shape moments of arcs, the rest is for
higher order moments and have less weight in the modeling.\\
A galaxy at the eastern part of A02 should weakly perturb its
location and shape. This galaxy was introduced in previous models but turns
out to have negligible consequences.
Indeed, only upper limits on its mass ($\sigma_v \lesssim 150\,km/s$) arise
when modeling. Its introduction appears marginally relevant for the study
and is ignored hereafter although its effect is shown on \reffig{genima_fig}.

\subsection{Strong lensing models without the fifth image}\label{model_1c_subsec}
\subsubsection{Dark matter density profiles}\label{dm_profiles_subsub}
We model the dark matter halo with two different density profiles.
In order to focus on the main differences between isothermal and NFW 
profiles, we keep the models as simple as possible and do not include 
peculiar features, like cluster galaxy perturbations.
The center of potential is allowed to move slightly within 2 arc-sec around
the cD of the cD galaxy. No prior assumptions are made about the ellipticity
and the orientation of the dark matter halo relative to the light nor to
the X-rays isophotes. 

The first profile is an elliptical isothermal distribution with
core radius of the form,
\be{rho_si_eq}
\rho(r)=\frac{\rho_c}{1+(r/r_c)^2}\quad {\rm ~in~spherical~approx.}
\ee
which is projected in,
\be{kappa_si_eq}
\kappa(r,\theta)=\frac{b}{\sqrt{\xi^2+r_c^2}},\quad\xi=r\sqrt{1-\epsilon\cos(2(\theta-\theta_0))}.
\ee

The core radius $r_c$, scale parameter $b=r_c+\sqrt{R_e^2+r_c^2}$, ellipticity
 $\epsilon$ and position angle $\theta_0$ are free parameters. $R_e$ is the
Einstein radius and $b$ is related to the cluster velocity dispersion by
\be{b_sigma_eq}
b=\left(\frac{\sigma_v}{186.5\,km.s^{-1}}\right)^2\frac{D_{ds}}{D_{s}}\quad{\rm arc-sec.}
\ee

The second profile is an elliptical NFW mass distribution. The 3D profile
has the form
\be{rho_nfw_eq}
\rho(r)=\frac{\rho_c\delta_c}{(r/r_s)(1+r/r_s)^2}\;\;\;\;\;\text{in spherical approx.}
\ee
where  $r_s$ is a scale radius, $\rho_c$ is the critical density of the
 universe at the redshift of the lens, and $\delta_c$ a concentration
parameter related to the ratio $c=r_{\rm 200}/r_s$ by
\be{concent_eq}
\delta_c=\frac{200}{3}\frac{c^3}{\ln(1+c)-c/(1+c)}.
\ee
The convergence $\kappa$ writes
\be{kappa_nfw_eq}
\kappa(r,\theta)=2\kappa_s\frac{1-F(x)}{x^2-1}\;\;\text{with\ \ }x=\xi/r_s\;,
\ee
where $\xi$ has the same meaning than before, 
$\kappa_s = \rho_c \delta_c r_s / \Sigma_{\rm crit}$ and
\be{fnfw_eq}
F(x)=\left\{\begin{array}{cl}
\frac{1}{\sqrt{x^2-1}}\tan^{-1}\sqrt{x^2-1}&\text{, for }x>1\\
\frac{1}{\sqrt{1-x^2}}\tanh^{-1}\sqrt{1-x^2}&\text{, for }x<1\\
1&\text{, for }x=1
\end{array}\right.
\ee

\subsubsection{Single halo best models}\label{best1c_models_subsub}
The inversion leads to two models that fit the strong lensing observations
equally well. They reproduce the multiply-imaged lens configuration of
both radial and tangential arcs. The NFW best fit model leads to a
$\chi_{\rm NFW}^2 = 2.42$ per degree of freedom and $\chi_{\rm IS}^2 = 3.0$
for the isothermal profile.\footnote{These values are higher than 1 but we
  remind that the models are significantly over-constrained.}
The final model parameters and errors bars are summarized in table
\ref{model_tab}. The centering of the dark matter halo relative to the
cD galaxy is discussed in Sect. \ref{center_cusp_subsec}.

\begin{table*}[ht]\centering
  \caption[]{Single halo models. The Einstein radius is the same in the best
    models for strong lensing : $R_E \simeq 50 \hmkpc$.
    For our two strong lensing models, the total mass inside this radius is
    $M(r<R_E) = (2.8\pm0.1)\times 10^{13}\,h^{-1}{\rm M}_\odot$. Errors due to
    uncertainties in arcs redshift are omitted. Also reported previous works
    results for comparison. Me93 refers to \citep{Mellier93}, Ha97 to
    \citep{Hammer97}, Mi95 to \citep{Miralda95} EF99 to \citep{Ettori99}
    and Al02 to \citep{Allen}. When known, the authors' values are
    recomputed in our adopted cosmology and with the 1.6 sources redshifts.
    For both papers, the center of potential location is assumed to match
    the center of cD or is not reported. Me93 core radii have been scaled in
    order to take into account the departs between their profile and an exact
    softened isothermal sphere. As well, Ha97 find a slope $\beta \approx 0.85$
    instead we have only considered models with $\beta = 1$ (see Eq.
    \eqref{hamprof_eq}). Al02 uses a NFW profile and only gives the scale radius
    value but we report on the same line our own measured values for ellipticity
    and position angle from Chandra X-ray brightness.
    The third column $r_c \backslash r_s$ corresponds either to the scale
    radius either to the core radius.
    Here, we convert all the position angles in a common definition, which
    is clockwise from North to East. The original paper do not report angles in
    the same frame but we made the correction except for Ha97 for which we do not know
    what is the reference. But in any case, the position angle is so constrained that
    these authors must have found a similar orientation as the other ones. Our
    definition is more valuable and self-consistent between Chandra, ROSAT,
    VLT and HST data.
    Models labels with a S refer to purely strong lensing modeling whereas a W stands
    for purely weak lensing fits. 
    The last row (cD+DM) concerns the last family of profiles with a cD and dark
    matter halo components and which is simultaneously constrained by strong+weak
    lensing. In the first column, we report the permitted inner slope for
    generalized NFW profiles (see Sect. \ref{slope_subsub}).}
  \label{model_tab} \begin{scriptsize}
    \begin{tabular}{lcccccccccc}\hline\hline\\
      model  &$\sigma_v$&$r_c \backslash r_s$&$\kappa_s$&$r_{\rm 200}$&c&$M_{\rm 200}$&$\epsilon$&$PA$&$x_c$&$y_c$\\
      &(km/s)&$(\hmkpc)$&&$(\hmkpc)$&&$(10^{14}\,h^{-1}\,{\rm M}_\odot)$&&deg&arcsec&arcsec\\\\\hline\\
      S-NFW &-&$90^{+35}_{-25}$&$0.6^{+0.5}_{-0.3}$&$920^{+180}_{-80}$&$12.5^{+5}_{-6}$&$\sim 5.5$&$0.24^{+0.04}_{-0.07}$&$58\pm1$&$0.1\pm0.4$&$0.2\pm0.4$\\ \\
      W-NFW &-& $67^{+300}_{-24} $&$0.74_{-0.5}^{+1.6}$&$890^{+160}_{-130}$& $12_{-8}^{+12}$&-&-&-&-\\ \\\hline\\
      S-isoT &$1022^{+40}_{-30}$&$10.4\pm1.8$&-&$\sim 1000$&-&$\sim 11.$&$0.25\pm0.05$&$59\pm1$&$0.2\pm0.4$&$0.2\pm0.4$\\ \\
      W-isoT & $900 \pm 150$& $<45$&-&-&-&-&-&-&-\\ \\\hline\\
      Me93 &$\sim 1000$&$4.5-7$&-&-&-&-&$0.15-0.33$&$51-66$&-&-\\ \\
      Mi95 &$\sim 1200$&$\sim10$&-&-&-&-&$\sim 0.22$&$\sim 58$&-&-\\ \\
      Ha97 &$\sim 1100$&$5-10$&-&-&-&-&$0.18-0.23$&$?47.5\pm5?$&-&-\\ \\
      EF99 &$\sim 930$&$\sim90$&-&$r_{500}\sim540$&-&-&-&-&-&-\\ \\
      Al02 &-&$107-120$&-&-&-&-&$\sim 0.20$&$58\pm7$&-&-\\ \\\hline\\
      cD+DM &$0.8<\alpha<1.1$&$\sim 85$&$\sim 0.6$&-&-&-&$0.22\pm0.06$&$58\pm2$& 0! & 0!\\ \\\hline\hline
    \end{tabular}
  \end{scriptsize}
\end{table*}

The associated counter-image of the radial arc A1 (bright and thin structure)
corresponds only to a small part of A5 that is triply imaged. Besides, the
diffuse component A'1 can be associated to A6 only if the corresponding source
is at a lower redshift than arc A1-A5. This corroborates photometric
redshifts results and was previously mentioned by \citeauthor{Hammer97}
Here, we find the source redshift $z_{\rm S(A6-A'1)}$ to be $1.1-1.3$.

The velocity dispersion derived for the IS model is consistent with results
of \citet{Mellier93}. The core radius proposed by these authors is 
higher because of its different definition. They used a
pseudo-isothermal projected gravitational potential\footnote{leading to a
  convergence : $\kappa(r) = \frac{b}{r_c}\frac{2+x^2}{(1+x^2)^{3/2}}$ with
  $x=r/r_c$ to be compared to Eq. \eqref{kappa_si_eq}}; instead, we directly model
the cluster projected density profile. Nevertheless, to ensure the same
Einstein radius with the same central velocity dispersion between their
model and ours, the core radius they reported must be twice the one we found.
Thus, core radii are consistent.

\subsection{Mass profile of MS2137.3-2353 from weak lensing analysis:}\label{WL_subsub}
 Although the error bars are large, the concentration parameter found for
the best NFW model is about twice the expectations from numerical simulations
and from the current measurements done in other clusters, even those with strong
lensing features \citep[See e.g.][]{Hoekstra02}. Since clusters are believed to be
triaxial, it may happen that the major axis lies along the line of sight, increasing
atificially its concentration by projection effects \citep[See e.g.][]{Jing02}.
However, it is likely that our model also mix together the contributions to the
effective concentration of the cluster and of the central cD potentials.
This possibility can be tested by comparing the strong lensing model with the weak
lensing anlysis that only probes the radial mass profile at larger scale,
where the cD contribution is negligible.
On large scales, we used the  VLT images to build a weak lensing catalog of
background galaxies covering a $6.4'\times6.4'$ field of view. At the cluster
redshift, this corresponds to a  physical radius of $700\hmkpc$. The detailed
description of the catalog analysis, namely PSF anisotropy corrections and
detailed galaxy weighting scheme and selection, is beyond the scope of this work.
The method we used can be found in \citep{Athreya02}. Here, we only compare the
result of the strong lensing mass profile and the fit of the azimuthally
averaged shear on scales $100\,\hmkpc\lesssim r \lesssim 1\,{\rm Mpc}$.\\
The shear profile is determined by using a maximum-likelihood analysis,
based on a $\chi^2$ minimization:
\be{chi_wl_eq}
\chi^2 = \sum_{i=1}^{N} \frac{|e_i - g(\vec{r}_i,z_i)|^2}{\sigma_{e,i}^2},
\ee
where $e_i$ is the complex image ellipticity, $g$ the complex reduced shear,
$z_i$ the photo-$z$ and $\sigma_{e,i}$ the dispersion coming from both the
intrinsic unknown source ellipticity and the observational uncertainties
\citep[See {\sl e.g.}][]{Schneider00}. In addition, we also measured the
$\zeta$-statistic and $\zeta_c$-statistic densitometry:
\begin{eqnarray}\label{zeta}
  \nonumber\zeta(r;r_{\rm max})&=&\bar{\kappa}(r'<r)-\bar{\kappa}(r<r'<r_{\rm max})\\
  &=&\frac{2}{1-(r_{\rm max}/r)^2}\int_r^{r_{\rm max}} \left< \gamma_t(r) \right> {\rm d} \ln r,\\
  \nonumber\zeta_c(r;r_2,r_{max})&=&\bar{\kappa}(r'<r)-\bar{\kappa}(r_2<r'<r_{\rm max})\\
  \nonumber&=& \left[1-(r/r_{\rm max})^2\right] \zeta(r;r_{\rm max}) + \\
  &&\frac{1-(r_2/r_{\rm max})^2}{(r_{\rm max}/r_2)^2-1} \zeta(r_2;r_{\rm max}) 
\end{eqnarray}
where $\left<\gamma_t(r) \right>$ is the averaged tangential ellipticity
inside an annulus ($\pi r^2 \Sigma_{\rm crit} \bar{\kappa}(r)$ is the mass
enclosed within the radius $r$). In contrast with the $\zeta$ statistics,
$\zeta_c$ can be directly compared with  $\bar{\kappa}$ since they only differ
by a constant value that does not change with radius \citep{Cloweetal00}.
We used $r_2 = 517 \hmkpc$ and $r_{\rm max}=744 \hmkpc$.
\\
The scaling factor for the mass has been derived from the $UBVRIJK$ photometric
redshifts of sources. Background galaxies have been selected in the magnitude
range $I<24$  and cluster galaxies have been rejected using a photo-$z$ selection.
Moreover, we considered background galaxies with \zp $\,>0.4$ for which
the lensing signal is significant. The limiting magnitude was chosen
in order to compromise between the depth, which defines the galaxy number
density, and the need for a good estimate of the source redshift distribution.
Since our source population is similar to \citet{VanWear90}, we checked our
redshift histogram has the same shape\footnote{after subtraction of the
  cluster population} as their sample. Both samples turned out to
be similar, so we finally used their parameterized redshift distribution,
because it is based on a larger sample than ours. With this requirement,
the weak lensing signal directly makes a test on the reliability of
strong lensing models extrapolations beyond the Einstein radius.

\begin{figure*}[htbp]
 \centering
  \includegraphics[width=18cm]{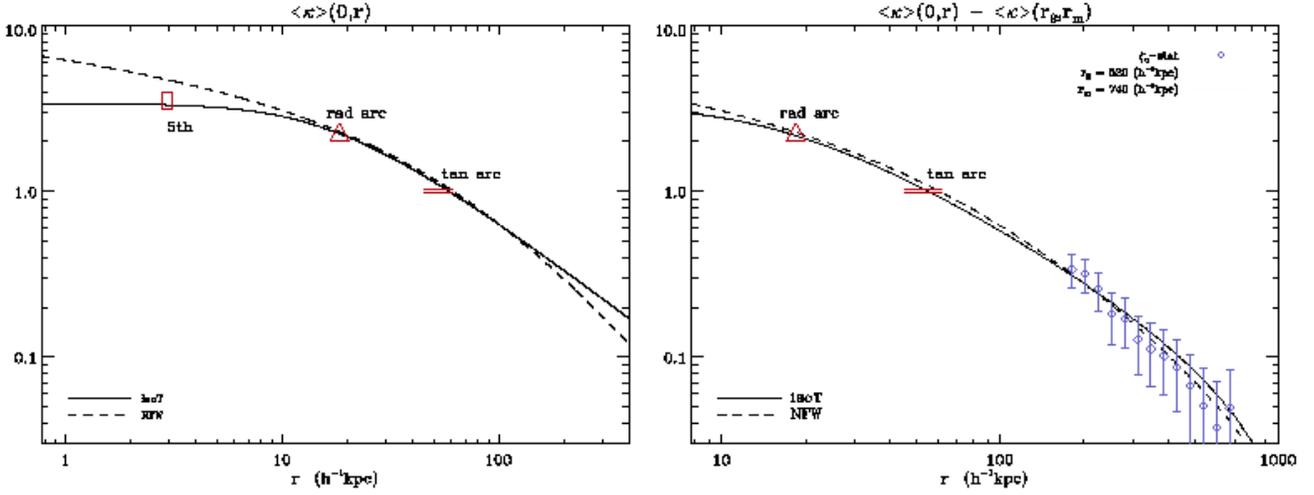}
  \caption{ Projected mean surface density. Solid:
    NFW best fit, dashed: IS. Notice how models match near the arcs locations.
    Between 2.5 and 28 arc-sec, the mean convergences $\bar{\kappa}=<\kappa>(0,r)$
    differ from each other by less than 3 percents. Faint discrepancies appear
    in the outer regions through weak lensing signal and at the very center.
    Constraints due to the fifth image candidate are detailed in Sect.
    \ref{slope_subsub}.
    The weak lensing data are deduced from the $\zeta_c$-statistic. }
  \label{kappa_fig}
\end{figure*}

Fig. \ref{kappa_fig} shows the radial mass profile of the best
IS and NFW models. The projected mass density has been averaged inside
circular annuli. As expected, the two best fits are quite similar between the
two critical radii. Discrepancies only appear in the innermost and outermost
regions. However, the shear profile derived from the VLT data fails to
disentangle the models built from strong lensing. Both are consistent with the
signal down to the virial radius $r_{\rm 200}\approx 1\,h^{-1}\,{\rm Mpc}$. Table
\ref{model_tab} lists the values of the best fit parameter set for the weak
lensing analysis. It is in good agreement with the inner strong
lensing  models, though the total encircled mass is smaller.
The constraints on the concentration parameter are weak and a broad
range of values are permitted. However, a low value similar as
expectations for clusters is still marginal and surprisingly the weak
lensing analysis also converges toward a rather larger concentration.
This discrepancy with cluster expectation values, even
when using together weak and strong lensing constraints, shows that
the global properties of the potential well are hard to reconcile
with a simple NFW mass profile. However, if the contribution of the
cD stellar mass profile strongly modifies the innermost mass
distribution of the cluster and significantly contaminates the
concentration parameter inward, our statement based on strong and
weak lensing models might be wrong. We therefore
single out the cD potential and add its contribution to the model
and we included the fifth image parameters in order
to probe the very center where the cD mass profile might play an
important role.

\subsection{The cD+DM mass profiles constrained with the fifth image.}\label{def_2c_subsec}
We now consider a two-component mass profile : an inner stellar component
attached to the cD galaxy and a cluster dark matter halo. The fifth
central image will contribute to constrain the innermost lens model,
whereas the external arcs and the weak lensing profile should constrain most
of the outer cluster halo.

\subsubsection{Centering the lens with the fifth image}\label{center_cusp_subsec}
\begin{figure}[htp]
    \centering \includegraphics[width=7cm]{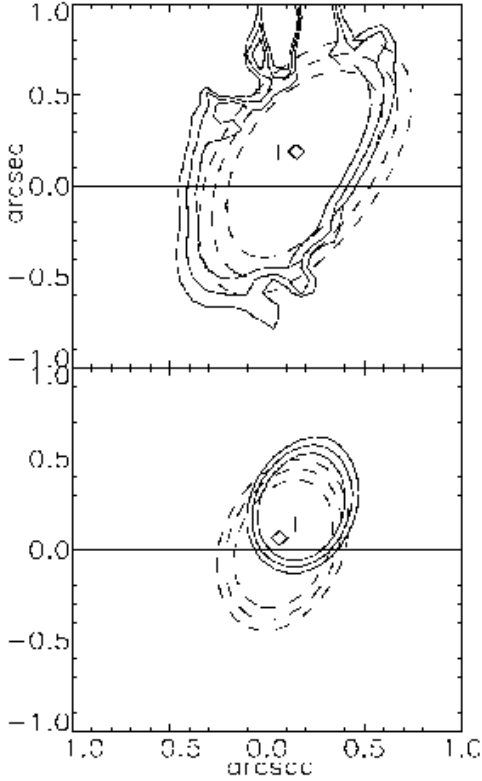}
    \caption{Permitted potential center region ($\Delta \chi^2/N =1,2,3$)
      relative to the cD center (0,0) coordinates.
      Solid (resp. dashed) contours refer to the NFW (resp. IS)
      model. The upper panel displays contours without the fifth image location
      knowledge and the lower using the 5th image candidate discussed in
      Sect. \ref{5th_cent_observ_subsec} as an additional constraint.
      The crosses (resp. diamonds) correspond to the $\chi^2$ minimum
      location for the NFW (resp. isothermal) modeling.
      Note the great enhancement induced and the significant shift in the NFW
      case compared to the upper panel. The deflection angle is merely
      too small to be consistent with the fifth image.}
    \label{center_and_5th_fig}
\end{figure}

Before introducing the stellar component, let us check the influence
of the fifth image knowledge on the centering of a single DM potential.
Fig. \ref{center_and_5th_fig}a shows the permitted area for the DM
potential center relative to the cD. The contours on the top
are the expectations for the IS and NFW models, if the fifth image
is not taken into account. The offset with respect to the cD centroid is
$0.22\arcsec$ West, but the contour ellipses are of size
$1.1\arcsec\times 1.6\arcsec$. Nevertheless, the assumption that the center
of cD galaxy coincides with the cluster center is consistent with the data.
When the fifth image is added (mainly its brightest knot),
the contours shrink by a factor of 2 in size, as shown in
\reffig{center_and_5th_fig}b, but still keep the central cD position inside,
with a small offset with respect to the cD light centroid
of $0.16\arcsec$ West for the IS model, and $0.18\arcsec$ West for the NFW
profile. The box sizes of permitted positions are much smaller ellipses
of about $0.6\arcsec\times 0.9\arcsec$. Since these error boxes are about
the size of the uncertainties of the cD centroid position (see Table
\ref{fifthtable}), in the following we will then assume the cD is centered 
on the cluster center. It is worth noticing that even with the significant
reduction of error bars provided by the fifth image, the residual uncertainty
on the centroid position of the lens may in principle permit to both IS and NFW
to fit the lensing data if we do not assume the cD center is not exactly on the
cluster center.

\subsubsection{Modeling together the stellar and DM mass profiles}\label{basic_2c_subsub}
The properties of the lens configuration (including the fifth image)
provide enough constraints to attempt a modeling which will probe clear
differences between observations and IS/NFW predictions.
The deflection and the magnification of the NFW model are smaller than for an
IS one. We expect the fifth image to show a difference of $0.2\arcsec$
in position and 0.75 in magnitude. The observations and IS/NFW predictions
reported in  table \ref{fifthtable} and Fig. \ref{center_fig} already
show a trend which supports a flat-core model against a cuspy NFW profile.\\
The following analysis uses together the fifth image 
properties, the weak lensing data and the giant arcs in
order to constrain the shape of the 
innermost mass profile. We also add the cD stellar
contribution to the overall mass because it
is no longer negligible at the very
center. Note that the generalized NFW models
expressed in  Eq. \eqref{nfw} has a free parameter $\alpha$.
Its projection is reported in  Eq. \eqref{kappa_cusp_proj}. In more details:
\begin{itemize}
\item The fifth image central coordinates are introduced because of their
  constraints on the central lens modeling. In fact, the
  brightest knot in the large arc A0 is required to correspond to
  the brightest detected spot at the center $r_5 \approx 0.9$.
\item The center of the cD is precisely the center of the potential well.
\item The dark matter halo is modeled by an elliptical mass distribution:
  a softened IS profile or a generalized-NFW profile, as expressed in
   Eq. \eqref{kappa_cusp_proj}. Hence, the IS profile
  has 4 free parameters, namely $r_c$, $\sigma_v$, $\epsilon$, $PA$ whereas
  the generalized-NFW models has 5 : $\alpha$, $r_s$, $\kappa_s$, $\epsilon$,
  and $PA$.
\item We model the stellar component with an Hernquist profile
  ($\rho(x)=x^{-1}(1+x)^{-3}$) of the projected form :
  \begin{equation}\label{hern_eq}
    \begin{split}
      \kappa_*(r) & = \frac{\kappa_{s,*}}{(y^2 - 1)^2}\left[-3 +
    (2+y^2) F(y)\right] \\
      \bar{\kappa}_*(r) & = 2 \kappa_{s,*} \frac{1-F(y)}{y^2 - 1}
    \end{split}
  \end{equation}
  with $F$ defined in Eq. \eqref{fnfw_eq}, $y=r/r_g$ and
  $r_g \approx 7.2\hmkpc$ a scale radius. $\kappa_{s,*}$ is related to
  the I band luminosity through the relation:
  $$\kappa_{s,*} = \frac{L_I}{2 \pi s^2 \Sigma_{crit}}
  \left(\frac{M}{L} \right)_I = 0.11 \Upsilon,$$
  where $\Upsilon \equiv M/L_I$. The stellar component is elliptical and
  has the central ellipticity $\epsilon_* = 0.15$ and orientation $PA_*=69^\circ$
  deduced from the light distribution in the I band. The cD stellar mass-to-light
  ratio $\Upsilon$ is the last new free parameter of the model.
\item The weak lensing $\chi^2$ term (Eq. \ref{chi_wl_eq}) is directly
  added to the purely strong lensing $\chi^2$ defined in Eq. \eqref{chisq_def_eq}.
\end{itemize}

\subsubsection{The inner slope of the DM halo}\label{slope_subsub}
The constraints provided by the fifth image can roughly be illustrated as
follows. If $r_5 \approx 0.9$ corresponds to a source position
$u\approx 1.85\arcsec$\footnote{Few  variations are observed when modeling
  with the NFW or the IS model. We also neglect ellipticity terms which have
  a weak importance near the center} deduced from the single component
outer arcs. The lens equation reads:
$u = r_5 \; \vert 1-\bar{\kappa}(r_5) \vert$
with the bending angle $\bar{\kappa}(r) = \tfrac{1}{r}\ddroit{\varphi}{r}$.
Hence, the averaged convergence within the fifth image candidate radius which
is plotted on \reffig{kappa_fig} is $$2.9\lesssim\bar{\kappa}(<r_{5})\lesssim3.7.$$

\begin{figure}[h]
  \centering \sidecaption
  \includegraphics[width=9cm]{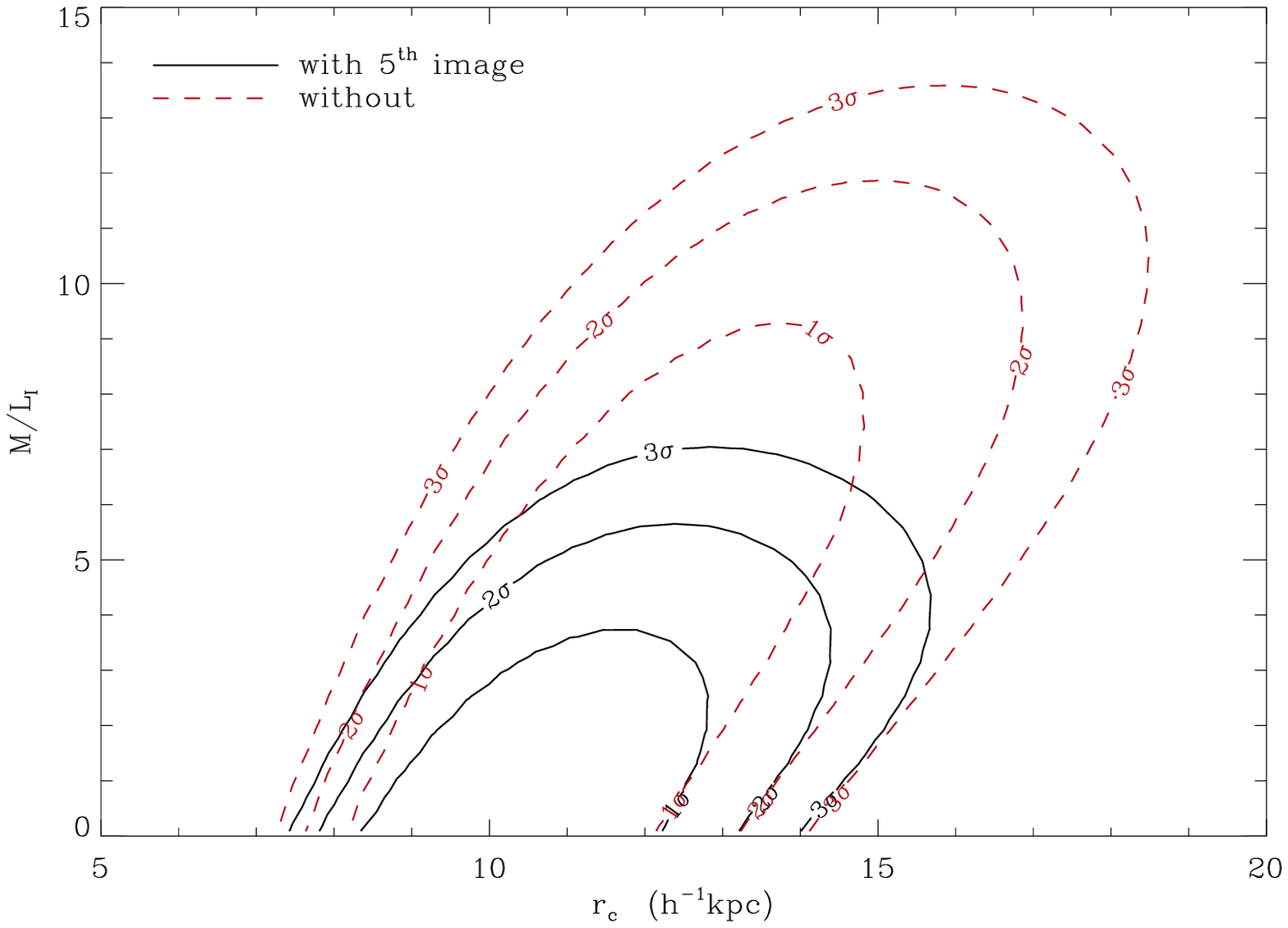}
  \includegraphics[width=9cm]{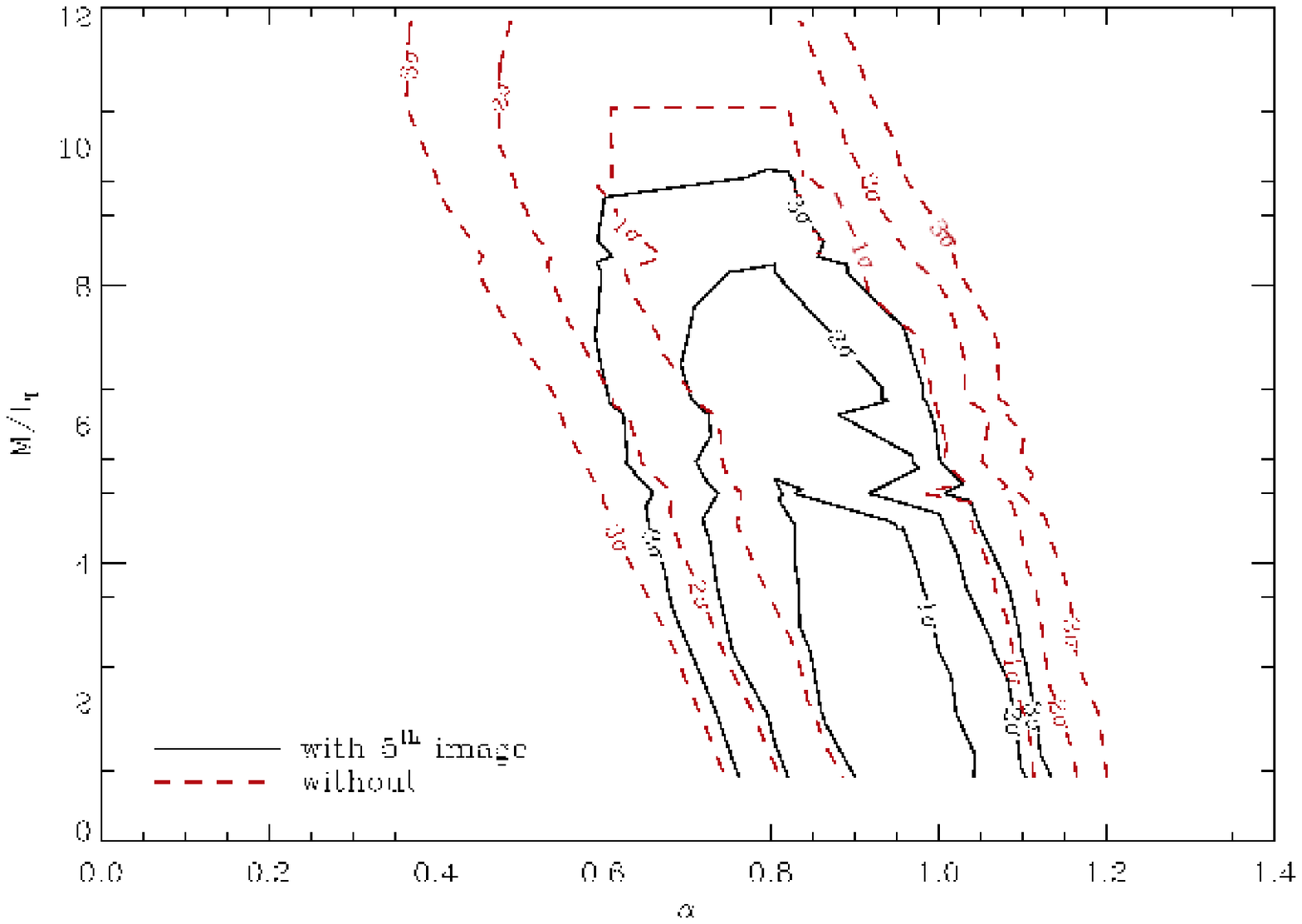}
  \caption{Two components $\chi^2$ contours. Upper panel : constraints on
    the couple $r_c-M/L_I$ for the IS model. Note the small modifications
    due to the introduction of the stellar component.
    Lower panel : constraints on the
    couple $\alpha-M/L_I$ for the pseudo-NFW family profiles.
    One can see an overlay of the same confidence regions when the fifth
    image position is known (solid) and when it is not (dashed).
    The inner slope $\alpha$ ranges between 0.8 and 1.1 for a reasonable
    mass of stars ($\Upsilon_I \leq 5$). The lower panel rejects very shallow
    profiles ($\alpha \leq 0.6$) whereas one can see on the upper one that
    flat cores provide good fits.}
  \label{full_max_lik_fig}
\end{figure}

It is now possible to test the overall permitted central contribution by
mapping the global $\chi^2$ in the  $r_c-M/L_I$ or $\alpha-M/L_I$ spaces,
after marginalization over the other parameters. For a reasonable
mass-to-light ratio $\approx 2-3$, one can see on \reffig{full_max_lik_fig}
that :
\begin{itemize}
\item Both IS and generalized-NFW models rule out $M/L_I$ of the cD
  stellar component larger than 9 at a 3-$\sigma$ confidence level.  Its
  value preferentially ranges within 1 to 5.
\item The softened IS profile still provides the best model.
  It is also consistent with the strong
  lensing data with few variations of parameters compared to the single
  component modeling. The introduction of the stellar component does not
  introduce large variations in the best fit parameters set compared to the
  single dark halo modeling.
\item The cuspy models have a narrow permitted range of slope which is
  centered around $\alpha=1$ of the NFW model.
  Note that while the position of the fifth image
  provides interesting boundaries on the cluster center position,
  it does not provide constraints
  on the slope $\alpha$. Including the fifth image only reduced
  the $\alpha$ upper limit by 10\% and does not changes its lower
  bound. For reasonable values of
  $M/L_I$, we find $0.8 \leq \alpha \leq 1.1$ (2-$\sigma$).
  This range excludes very low values of $\alpha$ and
  seems to contradict the fact that IS with flat core better fit
  the data than generalized-NFW models over the whole
  $1\hmkpc<r<600\hmkpc$ range,
  even for a large amount of stellar mass.
\end{itemize}

\section{Discussion}\label{discus_section}
\subsection{The radial mass profile of MS2137.3-2353}
The exceptional data set allowed us to constrain the density
profile over three orders of radius ranging from 2 to 700
$\hmkpc$. Despite the fact that weak lensing data do not cover a
wide enough range in order to reveal its full efficiency\footnote{the
  $\zeta$-statistic S/N ratio at radius $r$ goes like $\ln r_{\rm max}/r$ for an
  isothermal profile.}, we performed a self consistent modeling of the critical
strong and sub-critical weak parts of the lensing cluster MS2137.3-2353.
At the other side, it is worth noticing that the improvement provided by the
fifth image is  still under-exploited because of the poor resolution of its shape.
The location of its brightest spot only provides constraints on the
center position of the lens and on the overall enclosed mass
(by the way, revealing a degeneracy between $\Upsilon$ and $\alpha$).
A good knowledge of the magnification and shear would be able to break this
degeneracy by constraining second order moments of the fifth image probing
both convergence and shear inside 1 arc-sec radius.

\begin{figure}[ht]
  \centering
  \sidecaption
  \includegraphics[width=9cm]{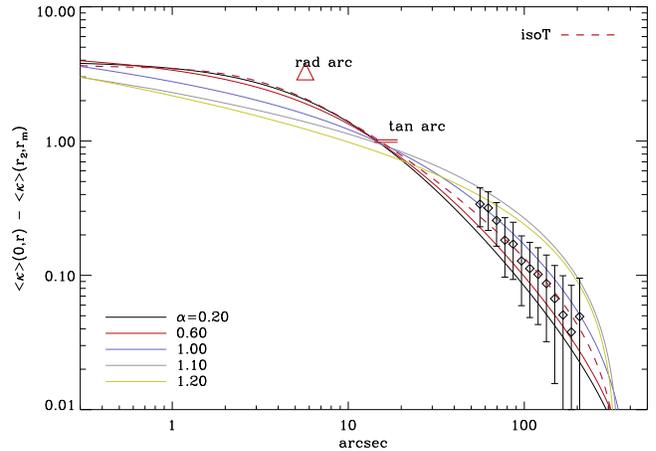}
  \caption{{\small
      Plot of various $\zeta_c$ curves taken along the relation
      plotted in blue (light gray) on \reffig{append_param_fig2}
      that give the observed critical lines. Also plotted the observed
      $\zeta_c(r)$ deduced from weak lensing.
      Only a small scatter around $\alpha=1$ is allowed.
      We also report the best fit isothermal profile (dashed).}}
  \label{append_param_fig1}
\end{figure}

\begin{figure}[ht]
  \centering   \sidecaption
  \includegraphics[width=9cm]{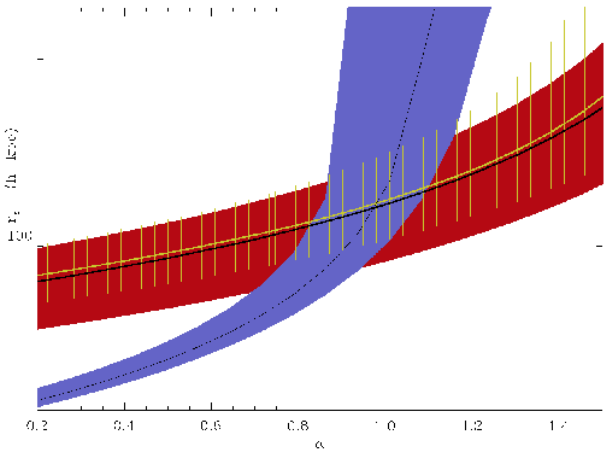}
  \caption{{\small Relations in the parameter space $\alpha-r_s$ deduced from
      the critical lines Eq. \eqref{system_anal_a} and Eq. \eqref{system_anal_b}
      (blue or light gray region), from the $\zeta-$statistic and
      Eq. \eqref{system_anal_a} (red or dark gray) and from the radial arc
      length relative to its counter-image (dashed yellow area).
      Errors take into account variations of ellipticity, of the mass-to-light
      ratio and observational uncertainties. The resulting permitted region
      is well consistent with what is found from the modeling. One can see
      that $0.7\lesssim\alpha\lesssim 1.1$. In other words, the halo density
      profile must be much shallower than the value 1.5 suggested by recent
      numerical simulations \citep{moore98,Ghigna00}. The permitted range for
      $r_s$ is restricted to small scale radii $r_s\sim 100\hmkpc $.}}
  \label{append_param_fig2}
\end{figure}

Nevertheless, the new constraints provide three levels of information concerning
the MS2137.3-2353 radial mass profile.
\begin{itemize}
\item[(i)] The simple use of the radial and tangential arcs systems without the
  fifth image neither the outer shear data cannot disentangle between NFW-like and
  isothermal profiles. A large family of cuspy and flat models are consistent with
  these data. (See Figs. \ref{append_param_fig1} and \ref{append_param_fig2}
  and appendix \ref{analytic_append})
\item[(ii)] The combined weak lensing and arcs (i) data tell us that either 
  isothermal profiles either NFW-like profile with $0.8\lesssim\alpha\lesssim 1.1$
  are permitted.
\item[(iii)] Actually, the fifth image knowledge favors flat cores 
  ($\chi^2_{IS} \sim 3.8$) but puts strong limits of the couple
  $\alpha-\Upsilon$ for NFW-like models ($\chi^2_{\alpha\sim 1}\sim 5.1$).
\end{itemize}
All together, the new constraints are in good agreement with
isothermal model with flat core and rule out generalized-NFW
models with slopes as steep as those proposed by \citep{moore98,Ghigna00}.
The slope range found for the generalized-NFW profiles can be easily
explained and correspond to expectations. The calculations detailed in
appendix \ref{analytic_append} (Figs. \ref{append_param_fig1} and
\ref{append_param_fig2}) show how the knowledge of the lensing configuration,
as derived from giant arcs and the shear field, bounds the free parameters
for cuspy-NFW profiles. The knowledge of the critical lines radii,
the weak lensing at intermediate scales as well as the length of the radial
arc are introduced in order to fix semi-analytically $r_s$, $\alpha$ and $k_s$.

A lower curvature for NFW-like profiles can explain the apparent paradox
discussed at the end of Sect. \ref{slope_subsub}. Flat softened isothermal
profiles are favored whereas low values of $\alpha$ are ruled-out. This trend
was reported by \citet{Miralda95} when he tried to fit the dark matter halo
with a density profile of the form
$$ \rho(x) \propto x^{-1}(1+x)^{-3/2}\quad {\rm or \ \ }\rho(x) \propto
x^{-1}(1+x)^{-1}.$$
Hence, a simple smooth modification from the inner slope $\alpha \ll 1$
to an asymptotic $\rho \propto r^{-3}$ outward cannot match the whole
lensing data. Sharper changes must occur at a small radius which
behaves as an effective core radius, leading to a high curvature close
to the radial arc. The scale radii derived from the marginalization of
\reffig{full_max_lik_fig} are very small and still scatter around
the $r_s \approx 90 \hmkpc$ value obtained from the single component NFW.
The sensitivity of models to radial curvature is clearly visible on
\reffig{append_param_fig1}.

The whole best fit generalized-NFW profiles show a high concentration
for the dark matter halo. This trend is confirmed by weak lensing up to
$700 \hmkpc$, in contrast with other weak lensing cluster analyses which
find smaller concentration than ours, but more consistent with
numerical CDM simulations. The role of stars does not change this
conclusion. So, if generalized-NFW models are acceptable, it is
important to confirm that in the case of MS2137.3-2353 they imply the
concentration to be stronger than numerical predictions. It is
therefore important to confirm these results by using a different method.
Recently, \citet{Sand} have reported comparable slope constraints using the
velocity dispersion of stars at the center of the cD and the positions
of critical lines. Conversely, any lensing model should be consistent with the
information on the kinetic of stars they measured, so it is necessary
to compare our predictions with their data. Nonetheless we plan to show
elsewhere (Gavazzi et al., in preparation) that the velocity dispersion
usually measured from the FWHM of absorption lines in the galaxy spectrum
no longer hold if the distribution function of stars is far from a
Maxwellian as mentioned in \citet{Miralda95}.

Finally, we checked that the introduction of galaxy halo perturbations under the form of massive
haloes attached to the surrounding cluster galaxies does not change our conclusions.
Such perturbations have poor consequences for the weak lensing results but are
likely affect slightly the fifth image location. We show in appendix
\ref{substruct_append} that a significant modification of the fifth image due
to galaxy halo perturbations implies to put a huge mass on each galaxy. Such an amount of mass
would destroy the quality of the arcs fit.

\subsection{Effects of non constant ellipticity and isodensity twist on
  the radial arc}\label{azimprof_subsec}
At the tangential arc radius ($r\sim 50\hmkpc$) we measure a robust offset
angle $\Delta\theta = 13^\circ\pm 4^\circ$ between the diffuse stellar
component and the DM potential orientation. This result is confirmed by
the Chandra X-rays isophotes contours as shown on \reffig{baryons_fig}.
Previous strong lensing modelings in the presence of important cD galaxies
never clearly established such a behavior because the uncertainties
of the models obtained with tangential arcs only were too loose for the
isopotential orientation.\\
However, for the nearby elliptical galaxy NGC720 \citet{Buote} and
\citet{Roman}\footnote{see references therein for a list of analogous objects}
studied such a misalignment between the light distribution
and the surrounding dark halo revealed by X-ray emissivity.
RK showed that the stellar misalignment can be explained by
a projection effect of triaxial distributions with aligned main axis
but different axis ratios.

Moreover all the best fit modelings show a tiny but robust remaining
azimuthal offset ($\lesssim 0.3 \arcsec$) between the modeled radial
arcs (A1 and A'1) and the position actually observed on the HST
image (highlighted on reconstructions of \reffig{genima_fig}).\\
We verified that it is not due to a bad estimation of the source position
since any small source displacement produces a large mismatch between the
counter-arc A5 and the model. This pure azimuthal offset led us to
investigate the possible effect of a variation of the ellipticity and
position angle of the projected potential close to the radial arc
radius. Such a trend is also favored by an increase of ellipticity on the
X-ray isophotes with radius.

If we had implemented the availability of using models with
a variation of orientation and ellipticity with radius in the inversion
software, we would have found that the orientation of the potential major
axis tends to the orientation of stars (see \reffig{baryons_fig}) when
looking further in. At the same time the potential becomes rounder.
We roughly checked this behavior by modeling the lens configuration with
two distinct (and discontinuous) concentric areas
(inside and above 8 arc-sec). The rays coming from the source plane and
giving rise to the outer arcs A0, A2, A4, A5 do not suffer any inner
variation of the potential symmetry (provided that the overall mass inclosed
in the Einstein radius remains the same). Hence the previous modelings
remain valid for the outer parts whereas the inner can be twisted and made
rounder in order to alleviate the offset problem. A small twist
$\Delta\,PA \simeq 8^\circ$ in the direction of stars gone with a smaller
ellipticity ($\epsilon \simeq 0.2$) turns out to suppress efficiently the
azimuthal offset near the center without affecting the external arcs.
This analysis is not exhaustive in the sense that maybe different
explanations can be found but its main virtue is to show that
high spatial resolution like HST imaging of numerous multiple arcs
makes a lens modeling so binding that it becomes possible to extract
much more information than the simple fit of
elliptical models. In addition with the hot ICM properties
\citep[see~e.g.~][]{Roman}, we could certainly start more detailed
studies of potential with twist effects and eventually start to probe
the triaxiality of dark matter halos if we can observe a large number
of multiple arc systems in clusters.
\\
These results strengthens the argument of \citet{Miralda02} upon which the
ellipticity of DM halos makes inconsistent the hypothesis of
self-interacting dark matter.

\section{Conclusion}\label{conclu_section}
By using strong and weak lensing analysis of HST and new VLT
data of MS2137.3-2353 we found important new features on the lensing
configuration:

The photometric redshifts or the radial and the tangential arcs
are both at $1.6\pm0.1$ in excellent agreement with the recent
spectroscopic observations of \citeauthor{Sand}.
\\
The extraction of the cD diffuse stellar light has permitted to
detect only one single object which turns out to be at the
expected position of the fifth image. Furthermore, its orientation,
its ellipticity, its signal-to-noise ratio and its morphology correspond
to those expected by the lens modeling. Unfortunately, the poor determination
of its shape properties hampers the use of its geometry as a local estimate
of the magnification matrix toward the center.
\\
Using the fifth image together with the weak lensing analysis of
VLT data, we then improved significantly the lens modeling. The
radial mass profile can then be probed over three orders in
distance. This additional constraint seems to favor isothermal
profiles with flat core or generalized-NFW profiles with
$0.8 \leq \alpha \leq 1.2$ without introducing the fifth image knowledge. 
When this constraint is added together with the prior motivated that cD center and
cluster halo center are the same, we favor flat core softened isothermal spheres. 
The position of the fifth image is in better agreement with an
isothermal model than an NFW mass profile.
In addition, it is worth noticing that the kinetics of stars should be analyzed
in details, considering a precise distribution function that may depart from
the commonly assumed Maxwellian.

We point out a misalignment between the diffuse stellar component
major axis and both the lens potential and the X-ray isophotes.
We argue it is produced by the triaxial shape of the mass components.
This extends the previous demonstration of the ellipticity of the
projected dark matter halo. This work is a first attempt to improve
strong lensing observables and modeling in order to probe both the central
DM cusp/core and the triaxiality of DM halos.

It would be essential to confirm the detection of the fifth image.
Fig. \ref{SED_fig} shows that the spectral energy distribution expected
for the fifth image is different than the old stars dominated cD emission.
We therefore expect the fifth image to show up on an optimal image
subtraction $U-\lambda J$ ($\lambda$ being optimized).
We attempted to use this technique on our present data but the poor
resolution ($\sim 0.6\arcsec$) on the U and J ground based images
prevent any significant enough detection. We conclude that only a high
resolution observation with the Space Telescope in UV-blue wavelengths
or in a peculiar emission-line is among the best constraints one could
envision in the future.

There is not yet evidence that similar studies as this work can be
carried out on other ideal lens configurations. The strength of
the diagnostic on the radial mass profile is however so critical
that we must apply this technique to a large sample in order to challenge
collision-less CDM predictions on a realistic number of clusters of galaxies
with eventually a test of the role of dominant central cD galaxies.
The simultaneous use of weak lensing data should be more relevant
for wider fields in order to check also a $r^{-3}$ fall-off
on the density profile predicted at large distance by CDM simulations.


\begin{acknowledgements}
  The authors would like to thank Jordi Miralda-Escud\'{e} for numerous and 
  fruitful discussions and comments on the content and the outlooks of that
  work. We also thank Peter Schneider, Tom Broadhurst and Avishay Dekel
  for useful comments. Special thanks to Stella Seitz who kindly provided
  their B and R VLT images of the cluster and for her comments on that work
  and to Chuck Keeton who made the last version of the {\tt lensmodel} software
  available. The authors also thank the referee for his 
  comments that help to clarify several points of the paper.
  This work was supported by the TMR Network
  ``Gravitational Lensing: New Constraints on Cosmology and the
  Distribution of Dark Matter'' of the EC under contract No. ERBFMRX-CT97-0172.
\end{acknowledgements}



\appendix 

\section{Table of identified dots.}\label{point_table_append_sec}
\begin{table*}[ht]
  \caption{X and Y coordinates in arc-sec for the 26 knots used in the
    modeling. Coordinates are referred to the cD center and are oriented as
    in \reffig{genima_fig}. The first part consists of the tangential arc
    system. The first 13 are 5 times imaged and the next 6 are triply imaged.
    The (*) denotes the brightest spot in the arcs and is the only point
    which are seriously observed in the fifth image. On the other side, the
    radial arc consists on triply images dots only. A1in (resp. A1out) is
    the inner (resp. outer) part of the composite radial arc.
    Both are imaged into the Eastern extremity of A5. In this later case,
    associations of mid points (x,y)=(8.3,-22.4) are somewhat ambiguous and
    are given less weight for the modeling.
    The tangential and radial systems were used simultaneously since we
    established they are at the same redshift $z_{\rm s,phot}\simeq1.6$.
    The system \{A'1-A6\} is found at a slightly lower photometric redshift and
    can only be compared with the modeling at a later stage.}
  \label{point_table_append} 
  \begin{scriptsize}
    \begin{tabular}{|r|rr|rr|rr|rr|rr|}\hline\hline
      ID &A02&&                A01&&         A2&&            A4&&         5th&  \\
      \hline
      1&-14.18 &  6.39  & -10.64 &11.67  & 12.06 & 5.24 & -2.65&-18.99  & 0.63 & 0.70\\
      2&-14.26 &  5.48  &  -8.65 &13.24  & 12.29 & 5.22 & -2.49&-18.79  & 0.65 & 0.77\\
      3&-14.28 &  5.74  &  -9.62 &12.58  & 12.14 & 5.38 & -2.38&-18.94  & 0.60 & 0.67\\
      4&-13.51 &  7.96  & -11.39 &10.92  & 11.98 & 5.48 & -2.46&-19.13  & 0.61 & 0.70\\
      5&-13.95 &  5.89  &  -8.35 &13.27  & 12.11 & 5.76 & -1.81&-18.99  & 0.61 & 0.63\\
      6&-14.00 &  5.22  &  -5.81 &14.54  & 12.33 & 6.12 & -0.78&-19.25  & 0.60 & 0.60\\
      7&-13.74 &  5.55  &  -5.32 &14.63  & 12.21 & 6.56 & -0.26&-19.20  & 0.54 & 0.61\\
      8&-13.71 &  5.39  &  -5.10 &14.80  & 12.38 & 6.45 & -0.08&-19.14  & 0.50 & 0.61\\
      9&-13.58  & 6.05  &  -5.73 &14.29  & 12.01 & 6.82 & -0.23&-19.32  & 0.51 & 0.65\\
      10&-12.65 &  7.78  &  -6.98 &13.45  & 11.52 & 7.40 & -0.24&-19.57  &(*)0.59& 0.66\\
      11&-12.43 &  8.16  &  -7.24 &13.26  & 11.50 & 7.22 & -0.71&-19.56  &0.57& 0.66\\
      12&-11.89 &  9.21  &  -9.08 &12.07  & 11.30 & 7.22 & -1.07&-19.57  & 0.60 & 0.73\\
      13&-11.13 & 10.42  & -11.60 & 9.91  & 11.48 & 6.99 & -1.24&-19.42  & 0.62 & 0.76\\
      14&&        &        &       &11.14  &7.06  &-1.48&-19.61   &0.62  &0.65\\
      15&&        &        &       &11.12  &6.84  &-1.88&-19.63   &0.65  &0.67\\
      16&&        &        &       &11.27  &6.02  &-3.10&-19.36   &0.70  &0.66\\
      17&&        &        &       &10.69  &5.90  &-3.92&-19.63   &0.79  &0.79\\
      18&&        &        &       &10.83  &5.62  &-4.31&-19.06   &0.73  &0.84\\
      19&&        &        &       &11.14  &5.07  &-4.76&-19.41   &0.77  &0.82\\\hline
    \end{tabular}
    \begin{tabular}{|l|rr|rr|rr|}\hline \hline
      ID&A5&&           A1in &&         A1out&\\
      \hline
      20&7.9 &-22.4&    -1.8 &  3.0  &   -4.1 &  5.4\\
      21&8.3 &-22.4&    -2.1 &  3.2  &   -3.9 &  5.1\\
      22&8.3 &-22.4&    -2.3 &  3.4  &   -3.6 &  4.6\\
      23&8.3 &-22.4&    -2.5 &  3.5  &   -3.6 &  4.3\\
      24&8.3 &-22.4&    -2.7 &  3.6  &   -3.2 &  4.1\\
      25&8.5 &-22.3&    -2.8 &  3.8  &   -3.0 &  4.0\\
      26&10.1 &-21.4&    -3.2 &  3.2  &   -3.8 &  3.8\\\hline
    \end{tabular}
  \end{scriptsize}
\end{table*}

\section{Analytical constraints on generalized-NFW profiles}\label{analytic_append}
 We consider the simplified case where the lens is described by an
elliptical density profile which has a small ellipticity $\varepsilon$
and a small stellar contribution. We neglect terms in the multi-polar
development higher than the quadrupole (first order in$\varepsilon$).

One can easily write a set of equations that the system must
verify : the critical lines locations, the lens equation relating
the radial arc A1 and its counter-image A5. One can also force the
model to fit the weak lensing constraints at large radius, say
$r_w = 160 \hmkpc$. The tangential line is known to pass by the
point
$$\vec{r}_t=[-11.25\arcsec;10.25\arcsec] \rightarrow r_t=15.2\arcsec,\;\varphi_t = 137.7^\circ,$$
the radial line to pass by the point
$$\vec{r}_r=[-3.16\arcsec;3.93\arcsec] \rightarrow r_r=5.0\arcsec,\;\varphi_r= 137.6^\circ,$$
the associated point in A5 is
$$\vec{r}_c=[8.3\arcsec;-22.4\arcsec] \rightarrow r_c=23.9\arcsec,\;\varphi_c=-69.5^\circ,$$
and the weak lensing $\zeta-$statistic constraint reads
\be{wl_anal_eq}
\bar{\kappa}(r_w) = 0.1 \pm 0.03.
\ee
In the following, subscripts t, r and c denote the values taken at position
$\vec{r}_t$, $\vec{r}_r$, $\vec{r}_c$ respectively. If we write the
magnification matrix as:
\be{magn_ap}
\mu^{-1} = \begin{pmatrix}
  1-\kappa+\gamma_1 & -\gamma_2 \\
  -\gamma_2 & 1-\kappa-\gamma_1
\end{pmatrix}
\ee
and the lens equation between radial arc A1 and its counter-image A5 position:
\be{lens_eq_ap}
\vec{u}=\vec{r}_r-\vec{\nabla}\phi(\vec{r_r})=\vec{r}_c-\vec{\nabla}\phi(\vec{r_c})
\ee

Reduced to the first terms in $\varepsilon$, the equation of the tangential
($\kappa+\gamma_1=1$), the radial ($\kappa-\gamma_1=1$) lines and the radial
relation between A1 and A5 respectively yield:
\begin{subequations}\label{system_anal}
  \begin{gather}
    \bar{\kappa}_t + [3\xi_t - \kappa_t] e_t  =  1 - S_t
    \label{system_anal_a} \\
    2\kappa_r -\bar{\kappa}_r - [3\xi_r - \kappa_r + r_r \kappa'(r_r)] e_r
    =  1 - S_r \label{system_anal_b}\\
    \bar{\kappa}_c + \eta \bar{\kappa}_r + e_c ( \xi_c -\kappa_c ) +
    e_r \eta ( \xi_r - \kappa_r ) = 1 + \eta - S_c \label{system_anal_c},
  \end{gather}
\end{subequations}
where $e_{i=t,r,c} = \varepsilon \cos(2(\varphi_i-\varphi_0))$,
$\kappa'=\ddroit{\kappa}{r}$, $\eta = r_r/r_c=0.21$ and
\be{xi_equ}
\xi(r) = \frac{2}{r^4} \int_0^r \der r'\,r'^3 \kappa(r')
\ee
\citep[fully detailed in][]{Miralda95}. The terms $S_{i=t,r,c}$
are the small corrections from the stellar
contribution. $S_t \approx \bar{\kappa}_*(r_t)$,
$S_r \approx 2\kappa_*(r_r) -\bar{\kappa}_*(r_r)$ and
$S_c \approx \eta \bar{\kappa}_*(r_r)+\bar{\kappa}_*(r_c)$ which are of a few
percents order and scale like the mass-to-light ratio $\Upsilon$.

If we now project a general 3D density profile ( \ref{nfw}) into:
\begin{equation}\label{kappa_cusp_proj}
  \begin{split}
    \kappa(r) &= 2 \kappa_s x^{1-\alpha} \lbrace (1+x)^{\alpha-3} +\\ &(3-\alpha)\int_0^1\negmedspace\der y\: (y+x)^{\alpha-4}(1-\sqrt{1-y^2})\rbrace\\
    \bar{\kappa}(r) &= 4 \kappa_s x^{1-\alpha} \lbrace\tfrac{1}{3-\alpha} \hypgeo(3-\alpha,3-\alpha,4-\alpha;-x)\\ & + \negmedspace \int_0^1\der y\: (y+x)^{\alpha-3}\frac{1-\sqrt{1-y^2}}{y}\rbrace
  \end{split}
\end{equation}
with $x=r/r_s$ and $\kappa_s\equiv\rho_s r_s /\Sigma_{\rm crit}$ we can
constrain all the parameters $r_s$, $\kappa_s $ and $\alpha$ for a given
ellipticity and a given mass-to-light ratio $\Upsilon$. We retrieve the NFW
profiles for $\alpha=1$. We also need to assume a position angle and an
ellipticity that we set equal to the values deduced from the modeling:
$\varphi_0=5^\circ\:,\:\varepsilon=0.24$. We analyzed departs from this value.

 In fact, we solve the set of Eq. \eqref{system_anal_a} and
Eq. \eqref{system_anal_b} for $r_s$ and $\kappa_s$ as a function of the inner
slope $\alpha$. Notice that the whole set of  Eq. \eqref{system_anal}
would in principle be sufficient for constraining exactly the triplet
$[\alpha,r_s,\kappa_s]$. Nevertheless, the radii inferred in these equations
are very similar and thus the solution suffers a high sensitivity to the
uncertainties on the values of $\vec{r}_r$, $\vec{r}_c$ and $\vec{r}_t$.

The numerical modeling deals with much more constraints than the
relations Eq. \eqref{system_anal} and Eq. \eqref{wl_anal_eq}. For example,
without the knowledge of the fifth image, the innermost constraint
given by the arcs on the density profile is the length of the
radial arc that extends down to 3 arc-sec from the center. Its
length depends on the source size which lies inside the caustic
and needs to be related to the shape of its counter-image A5. A
simple Taylor expansion of the lens equation around the radial
critical radius (where $\partial_{rr}\phi = 1$) relates the
half-length $\ell=1.8\arcsec$ of the radial arc to corresponding
source length $\der u$. This latter can be related to the size of
the arclet A5 ($\Delta A5_r=0.5\arcsec,\Delta A5_t=0.8\arcsec$)\footnote{Note
  that the ellipticity of the mass distribution implies that the
  magnification matrix is not diagonal. Hence, the radial and tangential
  lengths correspond to the radial arc length.} which is triply imaged:
\begin{subequations}\label{length_rad_arc}
  \begin{gather}
    \der u = \tfrac{\ell^2}{r_r}(1-\kappa_r - r_r \kappa'_r)\label{length_rad_ar_a}\\
    \nonumber\der u = (1-2\kappa_c+\bar{\kappa}_c + e_c (3\xi_c+r_c\kappa'_r-\kappa_c))\,\Delta A5_r\\
    + \varepsilon \sin(2(\varphi-\varphi_0)) (3\xi_c-2\kappa_c)\, \Delta A5_t\;.\label{length_rad_ar_b}
  \end{gather}
\end{subequations}

Eq. \eqref{length_rad_ar_a} uses the property
$\partial_{rrr}\phi(r_r) = (1-\kappa_r-r_r\kappa'_r)/r_r$ whereas
Eq. \eqref{length_rad_ar_b} uses
$\der u=(1-\kappa+\gamma_1) \der x_r -\gamma_2 \der x_t$.
 The equality between these two relations and the normalization at the Einstein
Radius (Eq. \eqref{system_anal_a}) constitute one more relation which is
plotted on \reffig{append_param_fig2}.

\section{Adding haloes of galaxies as substructures}\label{substruct_append}
\begin{figure}[ht]
  \centering  \sidecaption
  \includegraphics[width=9cm]{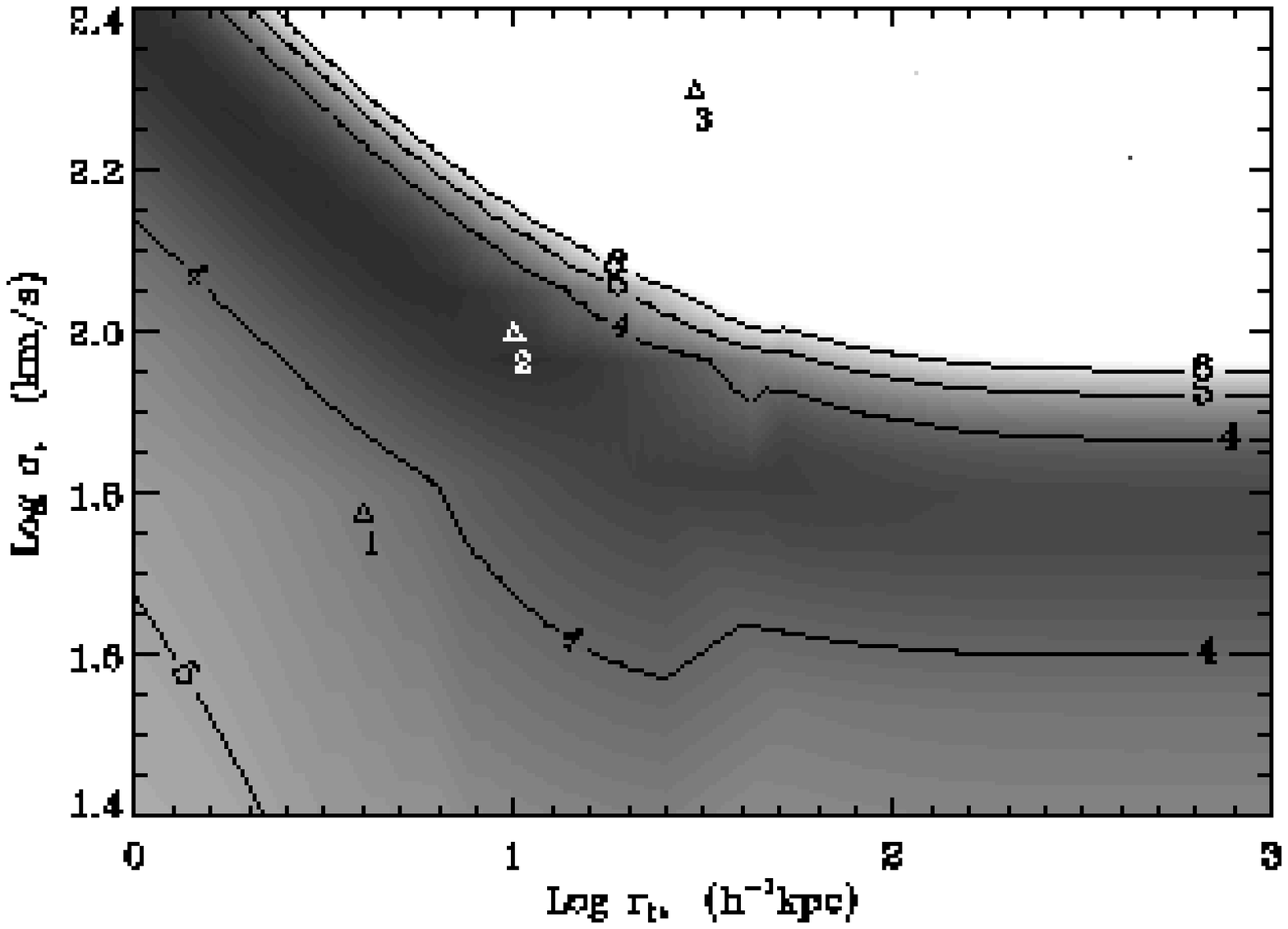}
  \includegraphics[width=11cm]{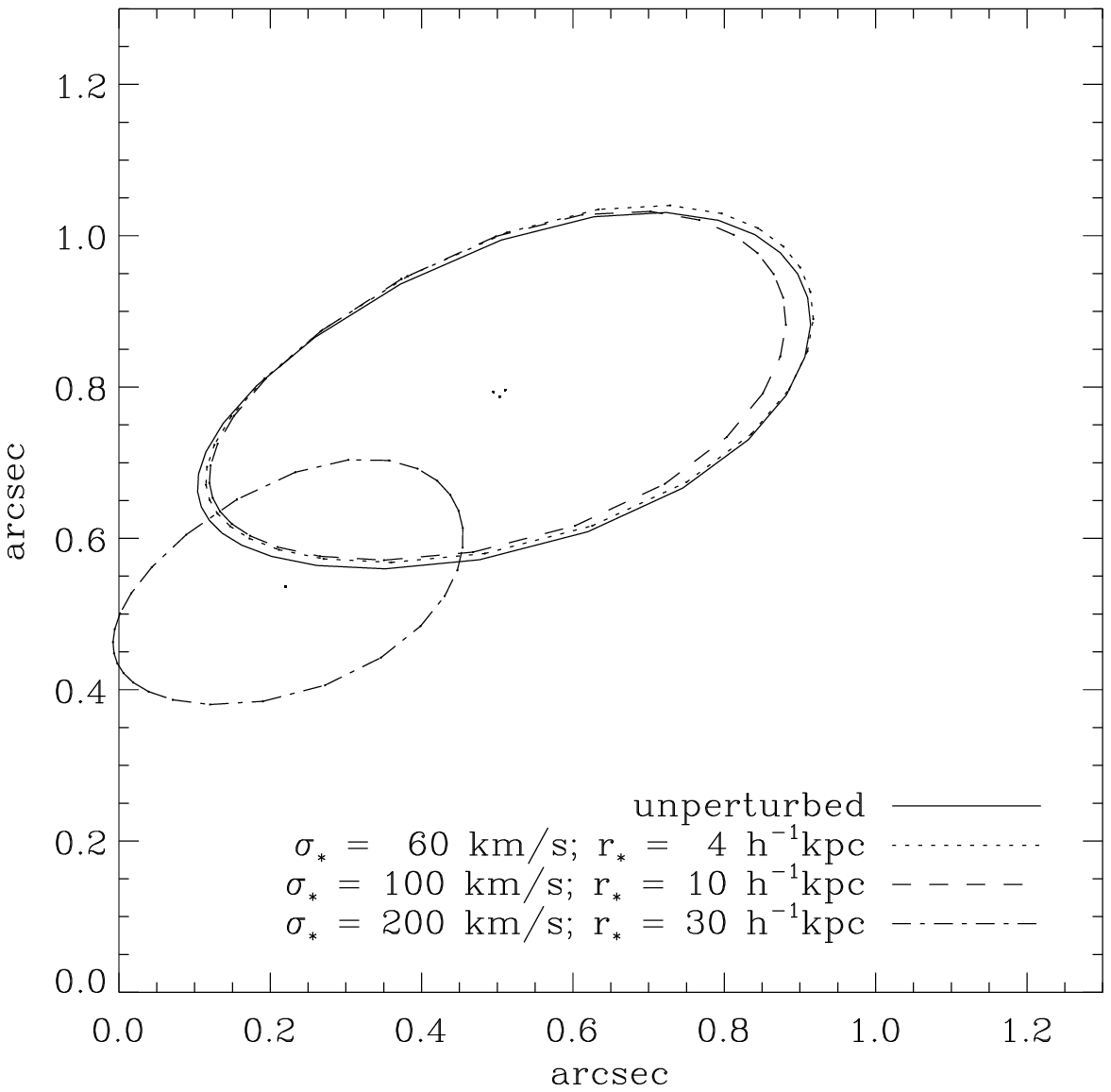}
  \caption{{\small Upper panel : $\chi^2/{\rm dof}$ contour plot for the couple
      $(r_{t*},\sigma_*)$. The ``macro'' model is the softened IS. Naturally,
      the $\chi^2$ minimum ($\sim 1.1$) is reduced as compared to the
      unperturbed previous model. The three peculiar couples 1, 2, 3 predict
      the fifth images plotted on the lower panel. The only noticeable effect of
      substructures on the fifth image occurs for models with a very high $\chi^2$
      model that provide a bad fit for the external arcs.}}
  \label{contours_perturb_fig}
\end{figure}

In order to demonstrate that adding galactic halos to the models  
has weak impact on our conclusions regarding the cluster mass profile,
we select the 9 brightest galaxies which are the closest from the
cluster center. Their I-band luminosities range between 
$0.17 < 10^{-10} {\rm L}_I/{\rm L}_{\sun} < 2.45 $ (the faintest has 
$L\approx0.17L_*$).
We adopt a Faber-Jackson scaling  to derive  their respective velocity dispersion.
Each galaxy halo density profile
is modeled by a truncated SIS with a cut-off radius $r_t$.
Without this truncation, halos perturbations are constant and propagate up to
the infinity. We adopt the scaling laws proposed by \citep{natarajan02}
to study the lensing cluster AC114:
\begin{subequations}\label{scaling_law}
  \begin{gather}
    \kappa_g(r) = \frac{b}{2} \left[\frac{1}{r}-\frac{1}{\sqrt{r^2+r_t^2}}\right]\\
    b = b_* \left(\frac{L}{L_*}\right)^{1/2},\quad\quad r_t = r_{t*} \left(\frac{L}{L_*}\right)^{1/2}
  \end{gather}
\end{subequations}
with $b$ related to $\sigma_v$ as in Eq. \eqref{b_sigma_eq}. $b_*$ and $r_{t*}$ are
two new free parameters. With this parameterization, perturbing galaxies have
an individual mass-to-light ratio that does not depends on their luminosity.
 In the following, we only show  the effect of perturbations
on the softened isothermal model, but 
   we found similar conclusions for the NFW model.
Fig. \ref{contours_perturb_fig} shows the $\chi^2$ contours for this new
couple of parameters after marginalization over the ``macro'' cluster model
parameters whereas the constraints are the ones used in Sect.
\ref{basic_2c_subsub} and include the fifth image brightest peak knowledge.

This modeling leads to a best fit $\chi^2/{\rm dof}$ much closer to 1.
It shows also that introducing galaxy halo perturbations
(in a way which is consistent
with the radial, tangential arcs and their counter-images ) still
predicts the  fifth image at the observed position. The NFW case is similar. 
We illustrate on 
Fig. \ref{contours_perturb_fig}b the effect on the fifth image equivalent
ellipse with the fiducial models referred as 1, 2 and 3 on
\ref{contours_perturb_fig}a and compare it to the unperturbed softened isothermal
model predictions. Galaxies haloes change the position and the shape 
of the fifth image only if they are so massive that they also 
damage significantly the external arcs image reconstruction. For instance,
it yields to a bad $\chi^2/{\rm dof} \sim 60$, in the third model case.


\begin{thebibliography}{!} \begin{small}

  \bibitem[Allen(1998)]{Allen98} Allen, S.~W.\ 1998, \mnras, 296, 392

  \bibitem[Allen et al.(2002)]{Allen} Allen, S.~W., Schmidt, R.~W., \&
    Fabian, A.~C.\ 2002, \mnras, 334, L11

  \bibitem[Arabadjis et al.(2002)]{Arabadjis02} Arabadjis,
    J.~S., Bautz, M.~W., \& Garmire, G.~P.\ 2002, \apj, 572, 66

  \bibitem[Athreya et al.(2002)]{Athreya02} Athreya, R.~M., Mellier, Y.,
    van Waerbeke, L., et al.\ 2002, \aap, 384, 743

  \bibitem[Bartelmann(1996)]{Bartel96} Bartelmann, M.\ 1996, \aap, 313, 697

  \bibitem[Bertin \& Arnouts(1996)]{Bertin96} Bertin, E.~\& Arnouts,
    S.\ 1996, \aaps, 117, 393

  \bibitem[Blumenthal et al.(1986)]{Blumenthal} Blumenthal, G.~R., Faber,
    S.~M., Flores, R., \& Primack, J.~R.\ 1986, \apj, 301, 27

  \bibitem[Bolzonella et al(2000)]{Bolzonellaetal00} Bolzonella, M.,
    Miralles, J.-M., \& Pell{\' o}, R.\ 2000, \aap, 363, 476

  \bibitem[Bode et al.(2001)]{bode01} Bode, P., Ostriker, J.-P., Turok, N.
    2000, \apj, 556, 93

  \bibitem[Bruzual \& Charlot(1993)]{Bruzual93} Bruzual, A., G.~\& Charlot,
    S.\ 1993, \apj, 405, 538

  \bibitem[Bullock et al.(2001a)]{BKW01} Bullock, J.~S., Kravtsov, A.~V.,
    \& Weinberg, D.~H.\ 2001, \apj, 548, 33 

  \bibitem[Bullock et al.(2001b)]{Bullock01} Bullock, J.~S., Kolatt, T.~S.,
    Sigad, Y., et al.\ 2001, \mnras, 321, 559

  \bibitem[Buote et al(2002)]{Buote} Buote, D.~A., Jeltema, T.~E.,
    Canizares, C.~R., \& Garmire, G.~P.\ 2002, \apj, 577, 183

  \bibitem[Burke(1981)]{Burke81} Burke, W.~L.\ 1981, \apjl, 244, L1

  \bibitem[Clowe et al.(2000)]{Cloweetal00} Clowe, D., Luppino, G.A.,
    Kaiser, N., Gioia, M.I. \ 2000, \apj, 539, 540

  \bibitem[Clowe \& Schneider(2001)]{Cloweetal01} Clowe, D., Schneider, P.\
    2001, \aap, 379, 384

  \bibitem[Dalal \& Kochanek(2002)]{Dalal02} Dalal, N.~\& Kochanek, C.~S.
    \ 2002, \apj, 572, 25

  \bibitem[Ettori \& Fabian(1999)]{Ettori99} Ettori, S.~\& Fabian, A.~C.\ 1999,
    \mnras, 305, 834

  \bibitem[Fort et al.(1992)]{Fort92} Fort, B., Le F\`{e}vre, O., Hammer, F.,
    \& Cailloux, M.\ 1992, \apjl, 399, L125

  \bibitem[Gavazzi(2002)]{gavazzi} Gavazzi, R.\ 2002, New Astronomy Review, 46, 783 

  \bibitem[Ghigna et al.(2000)]{Ghigna00} Ghigna, S., Moore, B., Governato,
    F., et al.\ 2000, \apj, 544, 616

  \bibitem[Gioia et al.(1990)]{Gioia90} Gioia, I.~M., Maccacaro, T.,
    Schild, R.~E., et al.\ 1990, \apjs, 72, 567.

  \bibitem[Haehnelt \& Kauffmann(2002)]{Haehnelt02} Haehnelt, M.~G.~\&
    Kauffmann, G.\ 2002, submitted to MNRAS., astro-ph/0208215

  \bibitem[Hoekstra et al.(2002)]{Hoekstra02} Hoekstra, H., Franx, M.,
    Kuijken, K., \& van Dokkum, P.~G.\ 2002, \mnras, 333, 911

  \bibitem[Hammer et al.(1997)]{Hammer97} Hammer, F., Gioia, I.~M.,
    Shaya, E.~J., et al.\ 1997, \apj, 491, 477

  \bibitem[Jing \& Suto(2002)]{Jing02} Jing, Y.~P.~\& Suto, Y.\ 
    2002, \apj, 574, 538 

  \bibitem[Keeton(2001a)]{Keea} Keeton, C.~R.\ 2001, \apj, 561, 46

  \bibitem[Keeton(2001b)]{Keeb} Keeton, C.~R.\ 2001, preprint, astro-ph/0102340

  \bibitem[Keeton(2001c)]{Keec} Keeton, C.~R.\ 2001, preprint, astro-ph/0111595

  \bibitem[Kelson et al.(2002)]{Kelson02} Kelson, D.~D., Zabludoff, A.~I.,
    Williams, K.~A., et al.\ 2002, \apj, 576, 720

  \bibitem[Klypin et al (1999)]{Klypin99} Klypin, A, Lravtsov, A, Valenzuela,
    O. \& Prada, F. 1999, \apj, 523, 32

  \bibitem[Kneib et al.(1993)]{Kneib93} Kneib, J.~P., Mellier, Y., Fort, B.,
    \& Mathez, G.\ 1993, \aap, 273, 367

  \bibitem[Kneib et al.(1996)]{Kneib96} Kneib, J.-P., Ellis, R.~S., Smail, I.,
    Couch, W.~J., \& Sharples, R.~M.\ 1996, \apj, 471, 643

 \bibitem[Maller \& Dekel(2002)]{Dekel02} Maller, A.~H.~\& Dekel, A.\ 2002,
    \mnras, 335, 487

  \bibitem[Mellier(1999)]{Mellier99} Mellier, Y. ARAA 37, 127.

  \bibitem[Mellier et al.(1993)]{Mellier93} Mellier, Y., Fort, B., \& Kneib,
    J.-P.\ 1993, \apj, 407, 33

  \bibitem[Merritt(1985a)]{MerrittA} Merritt, D.\ 1985, \aj, 90,
    102

  \bibitem[Merritt(1985b)]{MerrittB} Merritt, D.\ 1985, \mnras, 214, 25

  \bibitem[Metcalf \& Madau(2001)]{Metc01} Metcalf, R.~B.~\& Madau, P.\ 2001,
    \apj, 563, 9

  \bibitem[Milosavljevi{\' c} et al(2002)]{Merritt} Milosavljevi{\' c}, M.~,
    Merritt, D., Rest, A., \& van den Bosch, F.~C.\ 2002, \mnras, 331, L51

  \bibitem[Miralda-Escud\'{e}(1995)]{Miralda95} Miralda-Escud\'{e}, J.\ 1995,
    \apj, 438, 514

  \bibitem[Miralda-Escud\'{e}(2002)]{Miralda02} Miralda-Escud\'{e}, J.\ 2002,
    \apj, 564, 60

  \bibitem[Moore et al.(1998)]{moore98} Moore, B., Governato, F., Quinn, T.,
    Stadel, J. \& Lake, G.\ 1998, \apj, 499, L5

  \bibitem[Moore et al.(1999)]{Moore99} Moore, B., Ghigna, S., Governato, F.,
    et al.\ 1999, \apj, 524, L19

  \bibitem[Moore et al.(2000)]{moore00} Moore, B., Gelato, A.~J., Pearce,
    F.~R., Quilis, V.\ 2001, preprint, astro-ph/0002308

  \bibitem[Natarajan et al.Smail(2002)]{natarajan02} Natarajan, 
    P., Kneib, J., \& Smail, I.\ 2002, \apjl, 580, L11 

  \bibitem[Navarro et al.(1997)]{NFW} Navarro, J.~F., Frenk, C.~S.,
    \& White, S.~D.~M.\ 1997, \apj, 490, 493

  \bibitem[Navarro \& Steinmetz(2000)]{NavSt} Navarro, J.~F.~\& Steinmetz,
    M.\ 2000, \apj, 528, 607

  \bibitem[Osipkov(1979)]{Osipkov79} Osipkov, L.~P.\ 1979, Soviet
    Astronomy Letters, 5, 42

  \bibitem[Pell\'o et al.(2001)]{Pello01} Pell\'o, R., Bolzonella, M.,
    Campusano, L.~E., et al.\ 2001, Astrophysics and Space Science Supplement,
    277, 547

  \bibitem[Romanowsky \& Kochanek(1998)]{Roman} Romanowsky, A.~J.~\&
    Kochanek, C.~S.\ 1998, \apj, 493, 641

  \bibitem[Salucci \& Burkert(2000)]{Saluccietal00} Salucci, P., Burkert,
    A.\ 2000, \apj, 537, L9

  \bibitem[Sand et al.(2002)]{Sand} Sand, D.~J., Treu, T., \&
    Ellis, R.~S.\ 2002, \apjl, 574, L129

  \bibitem[Schneider et al.(1992)]{SEF92} Schneider, P.,
    Ehlers, J.,~\& Falco, E.~E.\ 1992, Gravitational Lenses,
    XIV.~Springer-Verlag. Also Astronomy and Astrophysics Library.

  \bibitem[Schneider et al(2000)]{Schneider00} Schneider, P., King, L.,
    \& Erben, T.\ 2000, \aap, 353, 41

  \bibitem[Smith et al.(2001)]{Smith01} Smith, G.~P., Kneib, J.-P., Ebeling,
    H., Czoske, O., \& Smail, I.\ 2001, \apj, 552, 493

  \bibitem[Sofue \& Rubin(2001)]{RotC} Sofue, Y.~\& Rubin, V.\ 2001,
    \araa, 39, 137
    
  \bibitem[Stocke et al.(1991)]{Stocke91} Stocke, J.~T., Morris, 
    S.~L., Gioia, I.~M., et al.\ 1991, \apjs, 76, 813 

  \bibitem[Spergel \& Steinhardt(2002)]{Spergel00} Spergel, D.~N., \&
    Steinhardt, P.~J., 2000, PRL, 84, 3760.

  \bibitem[Tyson et al.(1998)]{Tysonetal98} Tyson, J.~A., Kochanski, G.~P.,
    Dell'Antonio, I.~P.\ 1998, \apj, 498, L107

  \bibitem[Van Waerbeke et al.(2002)]{VanWear90} Van Waerbeke, L., Mellier,
    Y., Pell\'o, R., et al.\ 2002, \aap, 393, 369

  \bibitem[Verde \& Jimenez(2002)]{Verde02} Verde, L., Jimenez, R.\ 2002,
    submitted to MNRAS, astro-ph/0202283

  \bibitem[Willick \& Padmanabhan(2000)]{Willick00} Willick, J.~A.~\&
    Padmanabhan, N.\ 2000, preprint, astro-ph/0012253

  \bibitem[Wise \& McNamara(2001)]{wise01} Wise, M.~W.~\& McNamara,
    B.~R.\ 2001, Two Years of Science with Chandra, Abstract from the
    Symposium held in Washington~DC, September 2001, 217

  \end{small}
\end{thebibliography}
\end{document}